# Thermal transport from 1D- and 2D-confined nanostructures on silicon probed using coherent extreme UV light: General and predictive model yields new understanding

Albert Beardo[1,*,†], Joshua L. Knobloch[2,*,†], Lluc Sendra[1], Javier Bafaluy[1], Travis D. Frazer[2], Weilun Chao[3], Jorge N. Hernandez-Charpak[2], Henry C. Kapteyn[2], Begoña Abad[2], Margaret M. Murnane[2], F. Xavier Alvarez[1], Juan Camacho[1,*]

**Affiliations:**

[1]*Physics Department, Universitat Autònoma de Barcelona, 08193 Bellaterra, Catalonia, Spain*

[2]*Department of Physics, JILA, and STROBE NSF Science & Technology Center, University of Colorado and NIST, Boulder, CO 80309, USA*

[3]*Center for X-Ray Optics, Lawrence Berkeley National Laboratory, Berkeley, CA 94720, USA*

[†]*Albert Beardo and Joshua L. Knobloch contributed equally to this work.*

**Abstract:** Heat management is crucial in the design of nanoscale devices as the operating temperature determines their efficiency and lifetime. Past experimental and theoretical works exploring nanoscale heat transport in semiconductors addressed known deviations from Fourier's law modeling by including *effective* parameters, such as a size-dependent thermal conductivity. However, recent experiments have qualitatively shown new behavior that cannot be modeled in this way. Here, we combine advanced experiment and theory to show that the cooling of 1D- and 2D-confined nanoscale hot spots on silicon can be described using a general hydrodynamic heat transport model, contrary to previous understanding of heat flow in bulk silicon. We use a comprehensive set of extreme ultraviolet scatterometry measurements of non-diffusive transport from transiently heated nanolines and nanodots to validate and generalize our ab initio model, that does not need any geometry-dependent fitting parameters. This allows us to uncover the existence of two distinct timescales and heat transport mechanisms: an interface resistance regime that dominates on short time scales, and a hydrodynamic-like phonon transport regime that dominates on longer timescales. Moreover, our model can predict the full thermomechanical response on nanometer length scales and picosecond time scales for arbitrary geometries, providing an advanced practical tool for thermal management of nanoscale technologies. Furthermore, we derive analytical expressions for the transport timescales, valid for a subset of geometries, supplying a route for optimizing heat dissipation.

## I. INTRODUCTION

Advances in fabrication have scaled the characteristic dimensions of complex systems to the few nanometer range and even thinner. At these length scales, conventional macroscopic (bulk) models can fail to accurately describe nanoscale behavior, due to the dominance of interfaces and surfaces. Specifically, thermal transport from nanoscale heat sources on semiconductor substrates strongly deviates from bulk diffusive transport predictions. Experiments show that as the heat source size is reduced, the heat transport efficiency falls well below what is predicted by bulk diffusion, both for structured optical excitation[1–4] or for optically-excited nanostructured transducers[5–12]. Moreover, recent experiments have uncovered that both the size and spacing of periodic nanoheater arrays strongly influence thermal transport, resulting in new counter-intuitive behaviors[8,9]. However, there is still no consensus on the underlying physics—in large part because there is no comprehensive model to describe these new nanoscale thermal transport regimes. This precludes smart design for good thermal management in next-generation nanodevices.

New theoretical proposals based on truncated Levy flights[13], suppression of phonons[8,14] or relaxons[15] have explained certain aspects of non-diffusive thermal transport for specific geometries. Phonon hydrodynamics[16–24] has been also successfully used to explain thermal transport behavior on 2D materials[25], such as graphene[26], and even in bulk materials at very low temperatures[27]. As this behavior is known to occur when "normal" phonon scattering events (i.e. processes that conserve quasi-momentum) dominate over "resistive" ones, the existence of hydrodynamic transport in bulk semiconductors, like silicon, at room temperature has been explicitly discarded[28].

It is widely accepted that solving the Boltzmann transport equation with ab initio calculated parameters is the most precise way to describe the transport of phonons, which are the dominant heat carriers for semiconductor and dielectric materials[29–31]. However, several difficulties associated with this approach limit its use at a practical level. First, this equation is challenging to solve in general due to the complexity of phonon collisions. To overcome this challenge, the relaxation time approximation is often used to simplify the collision expression[13,32]. However, this approximation does not guarantee energy conservation, which can lead to invalid results[13,33]. Second, complex geometries are challenging because one must model how each phonon mode interacts with every boundary present. Finally, coupling this equation to other phenomena, such as thermoelectricity or thermoelasticity, exponentially increases the computational requirements. These challenges can prevent microscopic models, like the Boltzmann transport equation, from being directly compared to experimental data.

Because microscopic models cannot be applied in many complex geometries, experiments often use an intermediate layer, or mesoscopic models, to compare results and theory. Most mesoscopic models to date are based on Fourier's law of heat diffusion with the addition of phenomenological effective parameters. This approach fits effective parameters to experiments, and then formulates theoretical models to connect the fitted values to ab initio calculations. Recent works have used this effective Fourier model to analyze heat dissipation away from metallic nanostructures of varying size and spacing[5–9,11]. This can quantify the deviation from the diffusive prediction by fitting either an effective thermal boundary resistance between the transducer and substrate[7–9] or an effective thermal conductivity of the substrate[5,6,12,34–36]. These techniques have significantly advanced our understanding, making it possible to develop new experimental mean free path spectroscopy techniques[1], as well as uncovering new transport regimes dominated by the heat source spacing[7,8]. However, using Fourier's law as a mesoscopic model, even with effective parameters, can obscure the underlying physics and fails to predict thermal transport observed for all time and length scales[19,22]. Most importantly, this approach is difficult to generalize to arbitrary geometries or materials.

In this work, we present a comprehensive set of dynamic EUV scatterometry measurements of non-diffusive heat flow away from 1D- and 2D-confined nanostructures on bulk silicon. We use this data to validate and generalize the Kinetic Collective Model (KCM)[37,38], which is a mesoscopic model which uses a hydrodynamic-like heat transport equation[16] with ab initio parameters. Contrary to conventional understanding, we show that heat transport away from nanoscale sources on bulk silicon can be predicted by the hydrodynamic equation. This generalizes the hydrodynamic framework to situations where phonon momentum is conserved, which applies not only when normal collisions dominate, but in regions with size comparable to the average resistive phonon mean free path near heat sources and system boundaries[22,38,39]. We also experimentally observe for the first time that closely-spaced 2D-confined (nanodots) on a bulk silicon substrate cool faster than widely-spaced ones, and that this effect is larger in 2D-confined than in 1D-confined (nanoline) sources observed by previous works[8,9]. Moreover, we demonstrate that KCM both fully predicts the heat transport over a wide range of length-scales and time-scales from 1D- and 2D-confined heat sources on a silicon substrate—including the counter-intuitive behavior of the closely-spaced geometry—and captures the full thermomechanical response to the system, which is beyond the capabilities of microscopic models.

Our mesoscopic hydrodynamic model also provides insight into the fundamental transport behavior. KCM allows us to identify the time scales over which two different transport mechanisms are dominant: one characteristic time dominated by the thermal boundary resistance and another regime that is dominated by hydrodynamic heat transport. The latter mechanism is responsible for the slow thermal decay of small heat sources, and consequently its reduction is responsible for the increased dissipation of close-packed nanoheaters. Furthermore, we develop a two-box model, derived from the hydrodynamic equation, which provides a physical interpretation and specific expressions for the two characteristic dissipation mechanisms. We confirm these findings by comparing our models to both past 1D-confined and new 1D- and 2D-confined experimental data. We conclude that KCM—involving only a few parameters—provides a predictive description of the thermal and mechanical response in these complex systems with highly non-diffusive behavior and has specific advantages over the traditional effective Fourier model. This work thus represents a significant advance in both experimental and modeling capabilities opening the door to improved thermal management in iterative nanoscale device design, including possible routes to increase clock rates in nanoelectronics by surpassing what has been called the "thermal wall"[40,41].

## II. EXPERIMENTAL DESIGN AND THEORETICAL MODEL

We measure the heat dissipation away from 1D-confined periodic nanoline heat sources on a silicon substrate using dynamic EUV scatterometry similar to that of Refs.[7–9], but with significantly improved signal-to-noise ratio—by nearly two orders of magnitude. These improvements allow us to perform new measurements on 2D-confined periodic nanodot heat sources on a silicon substrate, which were fabricated under identical conditions as the nanoline arrays. Our time-resolved measurements use an ultrafast infrared pump laser pulse to rapidly excite thermal heating and expansion in the metallic structures. The resulting thermal and elastic surface deformation is monitored by measuring the change in diffraction efficiency of an ultrashort EUV probe pulse, as depicted in Figure 1 (see *Methods*). Using this technique, we observe the heat dissipation from nanodot arrays in general geometries without complex fabrication, for the first time, and of nanoline arrays down to 20nm in size (*L*) and 80nm in spacing (*P*).

To interpret the experimental data, we implement a mesoscopic model using KCM and a thermoelastic set of equations (see *Methods*). For heat transport, we use Fourier's law for the metal sources (which is dominated by electrons), and the Guyer and Krumhansl transport equation[16] for the substrate (silicon), which is the material where non-Fourier behavior is expected:

$$\tau \frac{dq}{dt} + q = -\kappa \nabla T + \ell^2 (\nabla^2 q + \alpha \nabla(\nabla \cdot q)), \qquad (1)$$

where $\kappa$ is the bulk thermal conductivity of substrate, $\tau$ the relaxation time of flux $q$, and $\ell$ the nonlocal length—that can be microscopically interpreted as a weighted average phonon mean free path. All these parameters are intrinsic properties of the substrate. Eq. (1) resembles the hydrodynamic Navier-Stokes equation of fluids; thus, we can build analogies between heat flow and fluid behavior. Fourier's law can be easily recovered from Eq. (1): when the experiment timescales are much larger than $\tau$, the first term can be neglected, and if spatial scales are much larger than $\ell$, the last term can be neglected. For sizes comparable to $\ell$, however, these viscous terms become important and capture the non-diffusive transport due to momentum conservation at the scale of the phonon mean free paths. In nanoscale regions near the heat sources, the momentum of emitted phonons is conserved due to the lack of resistive collisions. Hence, hydrodynamic effects can locally alter the heat transport even in semiconductors like silicon. In the limit where normal collisions dominate, Guyer and Krumhansl[16] found $\alpha=2$, however, we use $\alpha=1/3$—analogous to a fluid with zero volume viscosity—in agreement with more recent works[22,39,42,43]. To solve this equation, appropriate boundary conditions are implemented (see *Methods* and Supplementary Section 1). In addition, we require the thermal boundary resistance between the metal and the substrate. This boundary resistance is the only parameter in the model that cannot be derived from ab initio calculations, since it is highly dependent on the fabrication process rather than being an intrinsic material property. However, as all our nanograting arrays have been fabricated in identical conditions, we use the identical value of the thermal boundary resistance for the entire data set. Given the ab initio values for the other parameters, the model can be solved by using finite elements to determine the evolution of the displacement, the temperature, and the heat flux in the nanostructure and substrate[21].

The predictability of Eq. (1) has been recently validated in compact and holey films, and thermoreflectance experiments in silicon, with excellent agreement[21,22,44]. As discussed in Ref.[21], the applicability of the model with ab initio parameters ($\kappa$, $\ell$ and $\tau$) is restricted to geometries where edge effects produced by two different boundaries do not overlap, i.e. when boundaries are separated by a distance larger than $2\ell$. Here, the distance between heaters is $P - L$ (see Fig. 2). Thus, Eq. (1) is expected to be valid for nanostructure arrays satisfying $P - L > 2\ell$. We term experiments under this condition, where heaters are expected to behave independently, as effectively isolated heat sources, and those with $P - L < 2\ell$ as close-packed heat sources.

### III. RESULTS

We first study effectively isolated heat sources for both 1D-confined (nanolines) and 2D-confined (nanodots) of different sizes and periodicities. Figure 2 compares the experimental results on nanolines and nanodots with theoretical KCM solutions obtained using COMSOL. We compare both inertial solutions, that include elastic waves generated by the impulsive pump laser excitation, and quasi-static solutions without elastic waves to isolate the effects of the heat flow (see Supplementary Section 1). We use ab initio calculations to compute the intrinsic parameters of bulk silicon at $T$=300K: $\kappa$ = 145 W/mK, $\tau$ = 50ps, and $\ell$= 176nm[21,37]. For the thermal boundary resistance, which is an intrinsic property that depends only on the materials and the fabrication process, we use $R_1$=2.25 nKm$^2$/W for all nanostructure geometries (see Methods). The excellent agreement in Figure 2 between experiment and theory demonstrates an advance in modeling as the nanoline thermal decay has already been shown to be highly non-diffusive[8,9] and KCM—which is based on only a few key parameters—accurately predicts the thermal transport and elastic waves in both nanolines and nanodots without any geometry-dependent fit parameters, which is beyond the current capabilities of microscopic descriptions.

In the close-packed situation, $(P - L) < 2\ell$, nonlocal effects are expected to yield interaction between heaters, as phonons from a given source are able to reach neighboring sources before scattering. In this case, one does not expect Eq. (1) to be applicable since higher order derivatives should be included in the transport equation[35]. To keep the model as simple as possible, we propose that the effects of these higher order terms can be absorbed into a geometry-defined value $\ell_{\text{eff}}$, where Eq. (1) is still sufficient to describe the system. We propose the simplest expression that satisfies limiting cases: $\ell_{\text{eff}} = (P - L)/2 \ (< \ell)$. For this expression, when the period $P$ tends to the linewidth $L$, $\ell_{\text{eff}} \to 0$. In this limit, the grating tends to a line of infinite linewidth, thus viscous effects should vanish. In the other limit, if $(P - L) \to 2\ell$ we recover $\ell_{\text{eff}} \to \ell$ as constructed. Using this expression for $\ell_{\text{eff}}$, we compare KCM predictions with experimental results for close-packed nanoline(nanodot) heaters in Figure 1c(d). The model predicts that closely-spaced heat sources cool faster than widely-spaced ones, as uncovered in previous experiments[6,8,9]. We also experimentally demonstrate, for the first time, that this same counter-intuitive behavior observed in nanoline arrays is universal and manifests in nanodot arrays, since the L = 50nm with P = 200nm nanodot signal is relaxed at 800ps while L = 50nm with P = 400nm is not. The excellent agreement between the KCM prediction and experimental results for the close-packed cases shows that KCM can model this behavior with a simple expression for $\ell_{\text{eff}}$ (without fitting) while the other parameters used are the same used in the isolated cases. In summary, both nanoline and nanodot experiments can be predicted by the KCM using the intrinsic value $\ell$=176nm when sources are separated a distance larger than $2\ell$ (effectively isolated sources), and a geometry-defined effective value when distances are smaller (close-packed sources). This modification of $\ell$ for a specific situation allows us to retain both the predictive capability and simplicity of the model.

Using our model, we interpret the behavior of the effectively isolated sources from a hydrodynamic viewpoint and compare it to the close-packed sources. For effectively isolated sources, hydrodynamic effects become relevant when linewidth $L$ is on the same scale as the phonon mean free paths $\sim\ell$; thus, the non-diffusive terms in Eq. (1) reduce the heat flux, compared to Fourier's law, in agreement with experiments[5–9,12,44]. This phenomenon is analogous to a friction that arises from the large gradients in heat flux that impedes heat flow, referred to as a viscous resistance[38]. In other words, when linewidth $L$ is on the same scale as $\sim\ell$, there is not enough resistive phonon collisions to scatter the heat outward in all directions as diffusion assumes. Instead, the thermal energy is forced straight downward into the substrate over a distance related to $\sim\ell$ before enough resistive phonon collisions occur to dissipate energy in all directions, shown schematically in Figure 1. These hydrodynamic-like friction effects resulting from a lack of resistive collisions have been described in other formalisms albeit with different interpretations. For example, models using a phonon suppression function predict heat flow that is less efficient than Fourier's law when linewidth $L$ is on the same scale as $\sim\ell$, similar to our hydrodynamic model; however, this phenomenon is interpreted as a reduced number of carriers due to ballistically traveling phonons[14,34]. Additionally, models incorporating anisotropic behavior of thermal conductivity are parallel to the downward flux forcing predicted by our hydrodynamic model[10]. The viscous term in Eq. (1) naturally includes both heat flux reduction and apparent anisotropy observed by experiments. In Figure 3, we visualize these substrate regions where viscous effects are important (hydrodynamic regions) by converting results to a spatially-dependent effective thermal conductivity of silicon. Due to their proximity to the interface, if one tries to apply Fourier's law, hydrodynamic effects might be interpreted either as an increase of the thermal boundary resistance[8] or as a reduction of the thermal conductivity near the heater[5,32]. In the effectively isolated case with L=30nm and P=600nm, this region has a size of order $\ell \simeq$

200$nm$, while in the close-packed case (P=120nm), it is much smaller and of order $\ell_{\text{eff}} \simeq$ 50$nm$. Therefore, we hypothesize that the interaction of the nearby heat sources in the close-packed scenario reduces the non-local length, decreasing viscous effects, allowing the system to cool more efficiently than with isolated heaters. The microscopic description of this effect is the subject of future work.

To demonstrate the advantages of our hydrodynamic model over the traditional effective Fourier model with a best-fit boundary resistance, we compare the two theoretical predictions to experimental data for the isolated 250nm linewidth case in Figure 4. To emphasize the thermal decay of the system, we compare only the quasi-static calculations and data where the acoustic waves have been subtracted using the matrix pencil method (see Supplementary Section 3). Although an effective Fourier model can quantify the degree of the non-diffusive nature of the system, one finds that the best fit Fourier model fails to describe data at all times, as it overestimates the decay at the beginning and underestimates it at the end. In contrast, KCM predictions agree with data at all times. This plot indicates that the experimental results display two characteristic times: a fast one at short times and a slow one at longer times. These two different time scales are also apparent in the other nanostructure sizes shown in Figure 2. Therefore, as diffusive transport in these geometries contains only a single characteristic time scale, the effective Fourier model *cannot* capture the full nanostructure relaxation and misses the underlying physics, even with fitted intrinsic parameters (see Supplementary Section 2). We note that if the first ~500ps is excluded from the analysis—which is commonly done for experiments using visible probe beams due to the challenges of separating the contribution from hot electrons and thermal decay[5,6]—the presence of two distinct timescales cannot be observed (see Supplementary Section 2). However, our EUV probe is only sensitive to surface deformations (see Methods), and thus precisely captures two distinct timescales—a signature of non-Fourier heat transport.

A distinct advantage of KCM is that we can gain deeper insight into the two time scales of thermal relaxation by investigating the role played by hydrodynamics. To do this, we analytically solve the thermal equations in the heater and the substrate for the case $L < \ell$. In this range, hydrodynamic effects are dominant: the $q$ term in Eq. (1) can be neglected compared to the Laplacian term and the heat flux obeys the (linear) Navier-Stokes equation. The system of equations obtained is:

$$C_1 \frac{dT_1}{dt} = -\frac{T_1 - T_2}{R_1} \tag{2}$$

$$C_2 \frac{dT_2}{dt} = -\frac{T_2 - T_2^\ell}{R_2} + \frac{T_1 - T_2}{R_1}$$

where $T_1$ is the heater temperature, $T_2$ the average temperature of the substrate at the interface, and $T_2^\ell$ the average substrate temperature in the outer part of the hydrodynamic region, i.e. at a depth of order $\ell$ below the heater. $C_1 = c_h h$ denotes the heat capacity of the heater per unit surface, with $c_h$ and $h$ the specific heat and height of the heater, respectively. $C_2 = c_s L(1 + \alpha)/B$ is a heat capacity per unit surface characterizing the substrate, with $c_s$ the substrate specific heat, and $B$ a calculated geometric coefficient that for nanolines is 3.0. $R_1$ is the thermal boundary resistance between the metal and the substrate, and $R_2 = \frac{B\ell^2}{\kappa L}$ is a size-dependent thermal resistance due to viscous effects (details in Supplementary Section 2). At short times, $T_2^\ell$ is close to $T_2^\infty$ as heat has not reached this region and Eq. (2) becomes a linear system with a double-exponential decay:

$$T_1 - T_2^\infty = a_1 e^{-\frac{t}{\tau_1}} + a_2 e^{-\frac{t}{\tau_2}} \tag{3}$$

with $\tau_i$ and $a_i$ the characteristic times and weights, which are determined by $C_1, C_2, R_1$ and $R_2$. Therefore, KCM provides two characteristic times with specific expressions in terms of the physical properties of the system.

Equation (2) can be interpreted intuitively as a two-box model as seen in Figure 5. One box represents the heater, while the other box is a region of order $L$ in the substrate below the heater (referred to as the dam region). The thermal response of the system begins when the heater is filled with thermal energy from the laser pulse. At short times after the laser pulse, the heater releases the energy into the dam region, which retains the energy and rapidly increases in temperature. The initial rate of this energy transfer is dominated by the intrinsic thermal boundary resistance between the heater and substrate. At larger times, when the dam region has equilibrated with the heater, the dissipation of the thermal energy is dominated by the rate of energy transfer out of the dam region into the rest of the substrate. Therefore, the substrate plays two roles in the thermal response of the system: it acts both as an energy reservoir with heat capacity $C_2$ and as a thermal resistance $R_2$. The rate of energy transfer in these later times is controlled by the viscous resistance, i.e. hydrodynamic effects. The thermal relaxation of the heaters can be described by an equivalent circuit (Figure 5a) and illustrated by a fluid analog (Figure 5c). The predicted temperature evolution of the system as a function of time and position are shown in Figure 5b and Supplementary Section 2.

For small isolated sources, we find simple expressions for the characteristic times, namely $\tau_1 = R_1 C_{eq} = R_1 C_1 C_2/(C_1 + C_2)$, and $\tau_2 = (C_1 + C_2)R_2$. For nanolines of $L$=50 nm, these expressions yield $\tau_1 = 50\ ps$ and $\tau_2 = 1050\ ps$, thus $\tau_2$ is an order of magnitude larger than $\tau_1$. In this limit, $\tau_1$ depends on the thermal boundary resistance, while the viscous time scale $\tau_2$ does not depend on the thermal boundary resistance, but mainly on the nonlocal length $\ell$ and geometry:

$$\tau_2 = \frac{\ell^2 c_s}{\kappa}\left(1 + \alpha + B\frac{c_h}{c_s}\frac{h}{L}\right). \tag{4}$$

Therefore, for small isolated sources, KCM can provide simple analytical expressions for the two different time scales of the heat transfer, each one associated with a different resistive mechanism. This allows accurate experimental validation of the non-local length value for silicon at room temperature (a sensitivity analysis of various KCM parameters is provided in Supplementary Section 2). Additionally, the two-box model Eqs. (2) can also be applied to close-packed experiments by substituting $\ell$ by $\ell_{\text{eff}}$; however, the simple expression of Eq. (4) cannot be used in this case (see Supplementary Section 2).

Although the two-box model has been derived at small sizes, it also characterizes the non-Fourier behavior for all experimental sizes. To validate the intuition provided by the two-box model, we fit a double-exponential decay (Eq. (3)) to each of our experimental measurements, as shown in Figure 6a. We compare the fits of experiments to fits of numerical KCM simulations and the analytical two-box model in Figure 6b-d. We find that the experimental fit results agree well with both KCM numerical and analytical calculations.

Additionally, we confirm the existence of a short time scale ($\tau_1 \sim 100 \text{ps}$) which is dominated by the intrinsic thermal boundary resistance in Figure 6b and a longer time scale ($\tau_2 \sim 1 \text{ns}$) which is dominated by the hydrodynamic effects in Figure 6c. Figure 6c also displays the splitting of the decay times between effectively isolated and close-packed experiments, i.e. the increase in dissipation efficiency for close-packed heat sources. In Figure 6d, we plot the weight of the hydrodynamic dominated decay, $a_2$ in Eq. (3), which shows a transition from a primarily hydrodynamic decay for small heaters, to a decay ruled by the thermal boundary resistance at large sizes. This is expected as large sizes should converge to the Fourier prediction, which contains a single time scale. Therefore, the size-*dependent* effective boundary resistance extracted by the effective Fourier model in Refs.[8,9] can be re-interpreted as capturing the weighted average of the time-scales ($\tau_1, \tau_2$) generated by a size-*independent* boundary resistance and size-*dependent* localized hydrodynamic effects.

## IV. CONCLUSIONS

In conclusion, we have shown that by adding a hydrodynamic heat transport term, we can explain the thermal transport behavior of nanoscale metal-semiconductor samples with 1D- and 2D-confined heat source geometries over a large range of sizes, allowing a deeper insight of the physical behavior beyond effective Fourier's law. In addition, our formalism enables new predictive strategies to reduce the cooling time of nanoscale heaters. The mesoscopic character of the model and the use of intrinsic, geometry-independent parameters allows it to be easily extended to the complex architectures required by nanoscale technologies, where the lack of control on heat dissipation represents an important limitation for future developments.

**Materials and Methods:**

**EUV Dynamic Scatterometry Measurements:** The sample consists of metallic Ni nanostructure arrays fabricated on the surface of a silicon substrate using an e-beam lithography technique. The nanostructure arrays are 150x150µm² areas consisting of both periodic nanolines and nanodots with linewidths ranging from 1µm down to 20nm, periods ranging from 4µm down to 80nm, and average heights of 11.5nm. The linewidth and period of the nanoline/nanodot arrays is independently controlled in order to separate the effects of size and spacing. The dimensions of the various arrays are characterized using atomic force microscopy (see Supplementary Section 4). To launch dynamics in the sample, an ultrafast infrared (780nm wavelength, ~25fs pulse duration) pump beam is incident on the sample with ~20mJ/cm$^2$ fluence and ~275μm spot size. The pump light is preferentially absorbed by the metallic nanostructures which causes rapid heating followed by impulsive thermal expansion in the nanostructures. The coherent excitation of the periodic arrays launches acoustics waves that propagate along the surface of the silicon substrate. As the heated nanostructures cool down by thermal dissipation into the substrate, they relax back to their original profile. An ultrafast, short wavelength probe beam is generated by focusing an ultrafast infrared pulse into an Ar filled glass capillary. A quantum non-linear process called high harmonic generation converts a portion of the infrared light into a coherent short wavelength (~30nm) ultrashort pulse duration (~10fs) extreme ultraviolet (EUV) beam[45]. The short wavelength of the probe allows for exquisite picometer sensitivity to the surface displacement and allows for measurements of 10s nm nanostructures[46]. Moreover, these wavelengths interact with core electrons far from the Fermi surface which are not affected by small temperatures changes as the photon energies are far from resonances in nickel[46–48]. The probe beam is scattered from the nanostructure arrays at a set time delay, controlled by a mechanical delay stage, relative to the pump beam and captured on an EUV sensitive CCD camera. Images of the EUV scattering pattern with and without the pump beam are subtracted allowing us to observe the change in the diffraction pattern. By subtracting the change in intensity of the reflected EUV light from the change in intensity of the diffracted EUV light, we can compute the change in diffraction efficiency. This change in diffraction efficiency is monitored as a function of time delay between the pump and probe beams and can be directly related to the surface deformation of the sample.

**Thermoelastic modeling:** The microscopic expressions required for the ab initio calculation of the KCM parameters can be found in Refs.[37,38]. The same parameter values for silicon at room temperature have been used to model other experiments[21,22,38,44]

The temperature and the heat flux are obtained by solving the energy conservation equation along with the heat transport equation (Fourier's law for the heaters and Eq.1 for the substrate). The second order derivatives in the substrate transport equation (Eq. 1) require the inclusion of extra boundary conditions for the heat flux. A slip boundary condition relating the tangential heat flux in the substrate and its derivatives is imposed in the interfaces and in the silicon free surfaces (see Supplementary Section 1). In the free surfaces, thermal insulation is ensured by fixing normal component to zero. In the interfaces, we impose continuity of the heat flux normal component in the metal and in the substrate. Finally, we use a generalized boundary condition for the temperature jump in the interface including a Kapitza thermal boundary resistance term along with non-local terms[22]. Using ab initio calculations, we compute a lower bound for the thermal boundary resistance assuming diffusive phonon reflections and perfect contact area (see Supplementary Section 1). However, the nano-gratings fabrication process produces interface defects that increase the actual boundary resistance value. Therefore, a single correcting factor for the boundary parameters is required to predict the thermal decay of all the gratings (1D and 2D). The obtained correcting factor is fitted from the thermal decay of the largest experimentally available 1D grating (L=1µm) and hence does not depend on the model used. For large gratings, the hydrodynamic corrections do not play any role and we obtain the same boundary resistance correction using KCM or effective Fourier model. Specifically, we obtained a thermal boundary resistance value 3.1 times larger than the lower bound. This factor is similar to the one obtained in previous work for a similarly fabricated metal-semiconductor interface[22].

The thermal equations are coupled with the classical elastic equations to predict the surface deformation of the system in order to compute the resulting change in diffraction efficiency using numerical Fresnel propagation. Specifically, the stress tensor of the nickel and the silicon includes a linear thermal expansion term. Moreover, the thermo-elastic energy exchange term is included in the energy conservation equation. For heaters, we use nominal bulk nickel elastic properties. For the substrate, we use an anisotropic stress tensor accounting for the structural defects generated during the fabrication of the nano-gratings on the substrate top surface[49].

All the parameter values used and a detailed explanation of the thermoelastic equations and the boundary conditions can be found in Supplementary Section 1.

**Comparison between experiment and model predictions:** Since KCM consists of a linear set of partial differential equations, the surface deformation and the predicted diffraction efficiency linearly depends on the amount of energy deposited in the heater by the laser pulse. In the simulations, a uniform energy density of 1W/m$^3$ with a duration of <2.5ps is introduced in the heater. To compare the model predictions and the experiments, the diffraction efficiency obtained in KCM inertial simulation is scaled by a factor to match the first experimental peak. This is equivalent to scaling the simulated energy density by this factor and this same scaling is used to normalize the quasi-static simulations. This procedure is also applied to the effective Fourier simulations in order to compare Fourier and experiments in Figure 4. Note that a slight correction factor has been added to scaling of the quasi-static simulation of L = 30nm with P = 120nm in Figure 2d due to a small numerical error in the first few picoseconds of this inertial simulation.

The quasi-static solutions are obtained by removing inertial elastic effects, i.e. dynamic equilibrium is imposed during all the simulation (see Supplementary Section 1). These solutions capture the deformations just due to thermal expansion and hence can be used to track the temperature evolution of the system (see ref.[9]). Note that the initial peak obtained in the quasi-static simulations is not observed in experiments because the system needs a finite time to expand.

**Double Exponential Fitting to Experimental and Numerical Data:** For Figure 6, we performed double-exponential fits to both the experimental data and the numerical simulations. The quasi-static numerical solutions can be easily fit to a double-exponential using non-linear least squares; however, due to the noise and inertial elastic effects, a double-exponential function with four free parameters is too unconstrained to reliably fit to the experimental data. Therefore, we constrain the number of free parameters in a single fit while still independently extracting the four parameters of the double-exponential. We achieve this by fitting the data in several different steps. We determine a cut time, $t_c$, to divide the experimental trace into two parts in time to separately fit the two exponentials. We define the cut time as the time when the ratio between the two exponentials is 1% and compute $t_c$ using the fit values from the numerical solution. We fit the experimental data for times $t > t_c$ to extract the longer decay time exponential; however, the functional form of the decay for large

$t$ is not purely single exponential. Because diffusive transport occurs far from the heat source at large $t$, the decay has a power law component super-imposed on the exponential. To mitigate the effects of this power law on the extraction of the decay constant, we fit an effective Fourier model—with only two free parameters of effective thermal boundary resistance ($R_{eff}$) and the overall normalization ($A_2^{fit}$)—for $t > t_c$ and truncate the fit at roughly two times the expected decay constant. We can then convert $R_{eff}$ to a decay time, $\tau_2^{fit}$, since $\tau_2^{fit} = R_{eff} c_h h$.

To extract the other exponential, we correctly set the overall normalization by accounting for the inertial elastic effects in the experimental data. To do this, we fit the KCM quasi-static simulation to the experimental data with the acoustics waves subtracted for times $t < t_c$, constraining the maxima of the inertial KCM simulation and experimental data to be within the experimental noise. The resulting maximum of the KCM quasi-static simulation, $A$, allows us to compute $A_1^{fit} = A - A_2^{fit}$. We can extract the final parameter by fitting a double-exponential, $A_1^{fit} e^{-t/\tau_1^{fit}} + A_2^{fit} e^{-t/\tau_2^{fit}}$, for $t < t_c$ with only one free parameter, $\tau_1^{fit}$. For Figure6b-d, we plot $\tau_1 = \tau_1^{fit}$, $\tau_2 = \tau_2^{fit}$, and $a_2 = A_2^{fit}/A$, respectively, for both the experimental data and the numerical solutions. The error bars on the experimental data are the standard deviation from multiple measurements (if there are no error bars, then only one measurement was included). As the amplitudes of the exponentials ($a_1, a_2$) decrease, the extracted corresponding decay times ($\tau_1, \tau_2$) becomes more inaccurate. This is partially responsible for the difference between numerical solutions and experimental data for $\tau_2$ at large $L$ and $\tau_1$ at small $L$. Additionally, inertial effects and noise affect the extracted values of $\tau_1$ as $L$ decreases (see Supplementary Section 5).

**Acknowledgements:** A.B., L.S., J.B., F.X.A. and J.C. acknowledge financial support by the Spanish Ministerio de Ciencia, Innovación y Universidades under Grant No. RTI2018-097876-B-C22 (MCIU/AEI/FEDER, UE). J.L.K., T.D.F., J.N.H.C., H.C.K., B.A., and M.M.M. gratefully acknowledge support from the STROBE National Science Foundation Science & Technology Center, Grant No. DMR-1548924, and a Gordon and Betty Moore Foundation EPiQS Award GBMF4538. W.C. gratefully acknowledges support by the University of Colorado, Boulder through the U.S. Department of Energy under Contract No.

DE-AC02-05CH11231. J.L.K. acknowledges support from an SRC Fellowship. **Author contributions:** These authors contributed equally: Albert Beardo and Joshua L. Knobloch. M.M.M. and H.C.K. conceived the experiment. J.L.K., T.D.F. and B.A. performed the measurements and analyzed the experiment data. W.C. fabricated the samples. A.B., L.S., J.B., F.X.A. and J.C. performed the theoretical modelling and analysis. A.B. carried out the Finite Element modelling and L.S. the ab initio calculations. A.B., J.L.K., M.M.M., F.X.A, and J.C. wrote the article. All authors contributed to the study, discussed the results, and commented on the manuscript. M.M.M., H.K., B.A., J.N.H.C., F.X.A, and J.C. supervised the work. **Competing interests:** The authors declare no competing interests.
## References:

1.  Minnich, A. J. *et al.* Thermal Conductivity Spectroscopy Technique to Measure Phonon Mean Free Paths. *Phys. Rev. Lett.* **107**, 095901 (2011).
2.  Johnson, J. A. *et al.* Direct Measurement of Room-Temperature Nondiffusive Thermal Transport Over Micron Distances in a Silicon Membrane. *Phys. Rev. Lett.* **110**, 025901 (2013).
3.  Koh, Y. K. & Cahill, D. G. Frequency dependence of the thermal conductivity of semiconductor alloys. *Phys. Rev. B* **76**, 075207 (2007).
4.  Regner, K. T. *et al.* Broadband phonon mean free path contributions to thermal conductivity measured using frequency domain thermoreflectance. (2013) doi:10.1038/ncomms2630.
5.  Hu, Y., Zeng, L., Minnich, A. J., Dresselhaus, M. S. & Chen, G. Spectral mapping of thermal conductivity through nanoscale ballistic transport. *Nat. Nanotechnol.* **10**, 701–706 (2015).
6.  Zeng, L. *et al.* Measuring Phonon Mean Free Path Distributions by Probing Quasiballistic Phonon Transport in Grating Nanostructures. *Sci. Rep.* **5**, 17131 (2015).
7.  Siemens, M. E. *et al.* Quasi-ballistic thermal transport from nanoscale interfaces observed using ultrafast coherent soft X-ray beams. *Nat. Mater.* **9**, 26–30 (2010).
8.  Hoogeboom-Pot, K. M. *et al.* A new regime of nanoscale thermal transport: Collective diffusion increases dissipation efficiency. *Proc. Natl. Acad. Sci.* **112**,

DE-AC02-05CH11231. J.L.K. acknowledges support from an SRC Fellowship. **Author contributions:** These authors contributed equally: Albert Beardo and Joshua L. Knobloch. M.M.M. and H.C.K. conceived the experiment. J.L.K., T.D.F. and B.A. performed the measurements and analyzed the experiment data. W.C. fabricated the samples. A.B., L.S., J.B., F.X.A. and J.C. performed the theoretical modelling and analysis. A.B. carried out the Finite Element modelling and L.S. the ab initio calculations. A.B., J.L.K., M.M.M., F.X.A, and J.C. wrote the article. All authors contributed to the study, discussed the results, and commented on the manuscript. M.M.M., H.K., B.A., J.N.H.C., F.X.A, and J.C. supervised the work. **Competing interests:** The authors declare no competing interests.

## References:

1.  Minnich, A. J. *et al.* Thermal Conductivity Spectroscopy Technique to Measure Phonon Mean Free Paths. *Phys. Rev. Lett.* **107**, 095901 (2011).
2.  Johnson, J. A. *et al.* Direct Measurement of Room-Temperature Nondiffusive Thermal Transport Over Micron Distances in a Silicon Membrane. *Phys. Rev. Lett.* **110**, 025901 (2013).
3.  Koh, Y. K. & Cahill, D. G. Frequency dependence of the thermal conductivity of semiconductor alloys. *Phys. Rev. B* **76**, 075207 (2007).
4.  Regner, K. T. *et al.* Broadband phonon mean free path contributions to thermal conductivity measured using frequency domain thermoreflectance. (2013) doi:10.1038/ncomms2630.
5.  Hu, Y., Zeng, L., Minnich, A. J., Dresselhaus, M. S. & Chen, G. Spectral mapping of thermal conductivity through nanoscale ballistic transport. *Nat. Nanotechnol.* **10**, 701–706 (2015).
6.  Zeng, L. *et al.* Measuring Phonon Mean Free Path Distributions by Probing Quasiballistic Phonon Transport in Grating Nanostructures. *Sci. Rep.* **5**, 17131 (2015).
7.  Siemens, M. E. *et al.* Quasi-ballistic thermal transport from nanoscale interfaces observed using ultrafast coherent soft X-ray beams. *Nat. Mater.* **9**, 26–30 (2010).
8.  Hoogeboom-Pot, K. M. *et al.* A new regime of nanoscale thermal transport: Collective diffusion increases dissipation efficiency. *Proc. Natl. Acad. Sci.* **112**,

4846–4851 (2015).

9. Frazer, T. D. *et al.* Engineering Nanoscale Thermal Transport: Size- and Spacing-Dependent Cooling of Nanostructures. *Phys. Rev. Appl.* **11**, 024042 (2019).
10. Wilson, R. B. & Cahill, D. G. Anisotropic failure of Fourier theory in time-domain thermoreflectance experiments. *Nat. Commun.* **5**, 5075 (2014).
11. Oyake, T., Sakata, M. & Shiomi, J. Nanoscale thermal conductivity spectroscopy by using gold nano-islands heat absorbers. *Appl. Phys. Lett.* **106**, 073102 (2015).
12. Chen, X., Hua, C., Zhang, H., Ravichandran, N. K. & Minnich, A. J. Quasiballistic Thermal Transport from Nanoscale Heaters and the Role of the Spatial Frequency. *Phys. Rev. Appl.* **10**, 054068 (2018).
13. Vermeersch, B., Carrete, J., Mingo, N. & Shakouri, A. Superdiffusive heat conduction in semiconductor alloys. I. Theoretical foundations. *Phys. Rev. B - Condens. Matter Mater. Phys.* **91**, 085202 (2015).
14. Collins, K. C. *et al.* Non-diffusive relaxation of a transient thermal grating analyzed with the Boltzmann transport equation. *J. Appl. Phys.* **114**, (2013).
15. Cepellotti, A. & Marzari, N. Thermal Transport in Crystals as a Kinetic Theory of Relaxons. *Phys. Rev. X* **6**, 041013 (2016).
16. Guyer, R. A. R. A. & Krumhansl, J. A. Solution of the Linearized Phonon Boltzmann Equation. *Phys. Rev.* **148**, 766–778 (1966).
17. Guo, Y. & Wang, M. Phonon hydrodynamics for nanoscale heat transport at ordinary temperatures. *Phys. Rev. B* **97**, 035421 (2018).
18. Qu, Z., Wang, D. & Ma, Y. Nondiffusive thermal transport and prediction of the breakdown of Fourier's law in nanograting experiments. *AIP Adv.* **7**, 015108 (2017).
19. Ziabari, A. *et al.* Full-field thermal imaging of quasiballistic crosstalk reduction in nanoscale devices. *Nat. Commun.* **9**, 255 (2018).
20. Ding, Z. *et al.* Phonon Hydrodynamic Heat Conduction and Knudsen Minimum in Graphite. *Nano Lett.* **18**, 638–649 (2018).
21. Beardo, A. *et al.* Hydrodynamic Heat Transport in Compact and Holey Silicon Thin Films. *Phys. Rev. Appl.* **11**, 034003 (2019).
22. Beardo, A. *et al.* Phonon hydrodynamics in frequency-domain thermoreflectance experiments. *Phys. Rev. B* **101**, 075303 (2020).

23. Ma, Y. A two-parameter nondiffusive heat conduction model for data analysis in pump-probe experiments. *J. Appl. Phys.* **116**, 243505 (2014).
24. Ma, Y. Hotspot Size-Dependent Thermal Boundary Conductance in Nondiffusive Heat Conduction. *J. Heat Transfer* **137**, (2015).
25. Cepellotti, A. *et al.* Phonon hydrodynamics in two-dimensional materials. *Nat. Commun.* **6**, 6400 (2015).
26. Lee, S., Broido, D., Esfarjani, K. & Chen, G. Hydrodynamic phonon transport in suspended graphene. *Nat. Commun.* **6**, 6290 (2015).
27. Izawa, K. *et al.* Observation of Poiseuille flow of phonons in black phosphorus. *Sci. Adv.* **4**, eaat3374 (2018).
28. Simoncelli, M., Marzari, N. & Cepellotti, A. Generalization of Fourier's Law into Viscous Heat Equations. *Phys. Rev. X* **10**, 11019 (2020).
29. Li, W., Carrete, J., A. Katcho, N. & Mingo, N. ShengBTE: A solver of the Boltzmann transport equation for phonons. *Comput. Phys. Commun.* **185**, 1747–1758 (2014).
30. Carrete, J. *et al.* almaBTE : A solver of the space–time dependent Boltzmann transport equation for phonons in structured materials. *Comput. Phys. Commun.* **220**, 351–362 (2017).
31. Broido, D. A., Malorny, M., Birner, G., Mingo, N. & Stewart, D. a. Intrinsic lattice thermal conductivity of semiconductors from first principles. *Appl. Phys. Lett.* **91**, 231922 (2007).
32. Hua, C. & Minnich, A. J. Heat dissipation in the quasiballistic regime studied using the Boltzmann equation in the spatial frequency domain. *Phys. Rev. B* **97**, 1–7 (2018).
33. Minnich, A. J., Chen, G., Mansoor, S. & Yilbas, B. S. Quasiballistic heat transfer studied using the frequency-dependent Boltzmann transport equation. *Phys. Rev. B* **84**, 235207 (2011).
34. Maznev, A. A. & Johnson, J. A. Onset of nondiffusive phonon transport in transient thermal grating decay. *Phys. Rev. B - Condens. Matter Mater. Phys.* **84**, 195206 (2011).
35. Alvarez, F. X. & Jou, D. Size and frequency dependence of effective thermal

conductivity in nanosystems. *J. Appl. Phys.* **103**, 094321 (2008).
36. Huberman, S. *et al.* Unifying first-principles theoretical predictions and experimental measurements of size effects in thermal transport in SiGe alloys. **1**, (2017).
37. Torres, P. *et al.* First principles kinetic-collective thermal conductivity of semiconductors. *Phys. Rev. B* **95**, 165407 (2017).
38. Torres, P. *et al.* Emergence of hydrodynamic heat transport in semiconductors at the nanoscale. *Phys. Rev. Mater.* **2**, 076001 (2018).
39. Guo, Y. & Wang, M. Phonon hydrodynamics for nanoscale heat transport at ordinary temperatures. *Phys. Rev. B* **97**, 035421 (2018).
40. TECHNOLOGY; Intel's Big Shift After Hitting Technical Wall. https://www.nytimes.com/2004/05/17/business/technology-intel-s-big-shift-after-hitting-technical-wall.html.
41. Waldrop, M. M. The chips are down for Moore's law. *Nature* **530**, 144–147 (2016).
42. Guo, Y., Jou, D. & Wang, M. Nonequilibrium thermodynamics of phonon hydrodynamic model for nanoscale heat transport. *Phys. Rev. B* **98**, 104304 (2018).
43. Mohammadzadeh, A. & Struchtrup, H. A moment model for phonon transport at room temperature. *Contin. Mech. Thermodyn.* **29**, 117–144 (2017).
44. Alajlouni, S. *et al.* Geometrical quasi-ballistic effects on thermal transport in nanostructured devices. *Nano Res.* **14**, 945–952 (2021).
45. Rundquist, A. *et al.* Phase-matched generation of coherent soft x-rays. *Science (80-. ).* **280**, 1412–1415 (1998).
46. Tobey, R. I. *et al.* Ultrafast extreme ultraviolet holography: dynamic monitoring of surface deformation. *Opt. Lett.* **32**, 286 (2007).
47. Bencivenga, F. *et al.* Nanoscale transient gratings excited and probed by extreme ultraviolet femtosecond pulses. *Sci. Adv.* **5**, eaaw5805 (2019).
48. Naumenko, D. *et al.* Thermoelasticity of Nanoscale Silicon Carbide Membranes Excited by Extreme Ultraviolet Transient Gratings: Implications for Mechanical and Thermal Management. *ACS Appl. Nano Mater.* **2**, 5132–5139 (2019).
49. Nardi, D. *et al.* Probing thermomechanics at the nanoscale: Impulsively excited pseudosurface acoustic waves in hypersonic phononic crystals. *Nano Lett.* **11**, 4126–4133 (2011).

**Figures and Tables:**

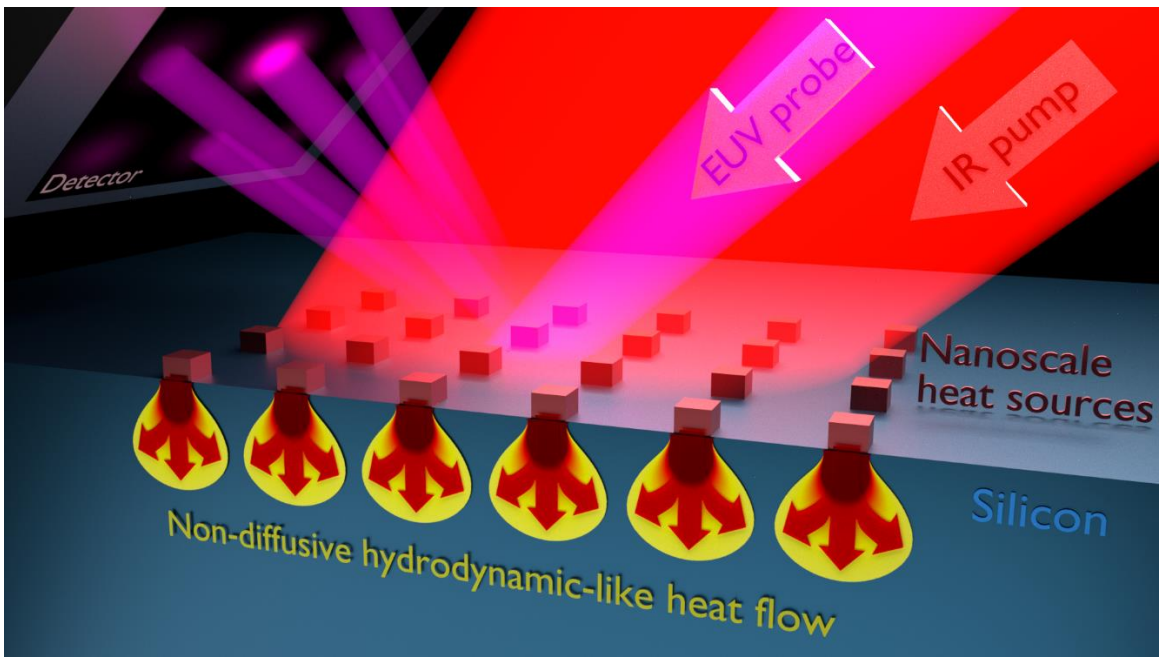

*Figure 1. Schematic of dynamic EUV scatterometry for probing non-diffusive hydrodynamic-like heat flow.* An ultrafast laser pulse rapidly heats the nanostructured transducers, which dissipate the thermal energy by transferring heat to the substrate. The heat flows away from the nanoscale heat sources following a non-diffusive hydrodynamic-like behavior, creating a 'balloon' shaped temperature profile. The resulting surface deformation of the heated nanostructures and substrate is measured via diffraction of an ultrafast Extreme Ultraviolet (EUV) probe pulse, after a controlled pump-probe time delay. The EUV pulse scatters from the periodic nanostructure arrays into a detector. We reduce the recorded scattering pattern into a single value of diffraction efficiency as a function of time delay between the pump and probe pulses, which precisely tracks the thermal and elastic dynamics in the sample.

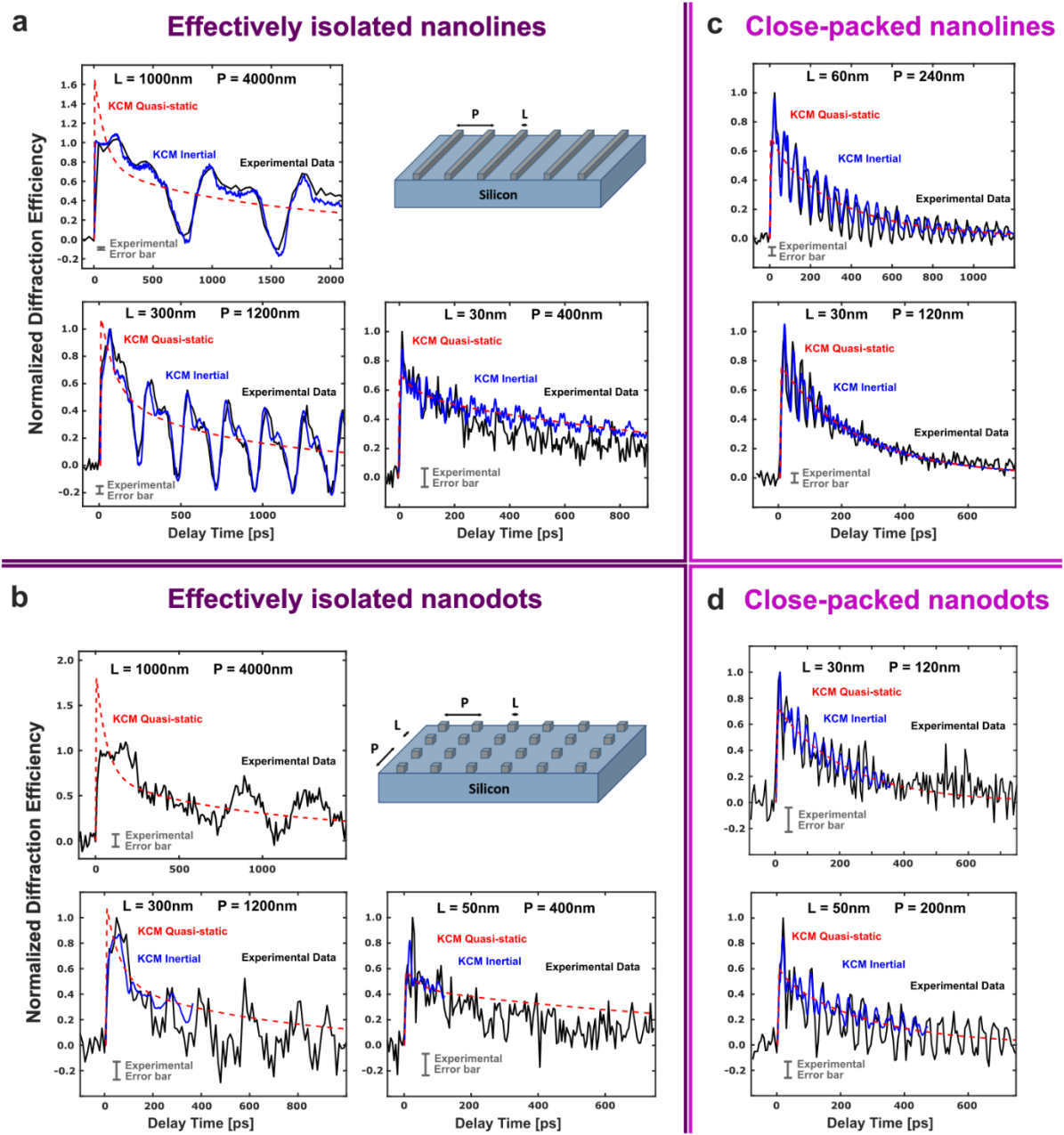

***Figure 2. Direct comparison between EUV scatterometry data and KCM modeling in 1 and 2D.*** *Experimental and theoretical normalized change in diffraction efficiency as a function of delay time for different sizes L and periods P for **(a)** effectively isolated, i.e. where $(P - L) > 2\ell$, nanolines (1D) and **(b)** nanodots (2D). Black lines denote experimental data where the error is represented by the gray bar. Blue lines indicate the inertial KCM predictions, and red lines denote the KCM quasi-static predictions which describe only the thermal transport without the contribution of oscillating elastic waves. Theoretical*

*predictions are computed using the same geometry-independent parameters for all nanostructure sizes and shapes. The theoretical curves are identically normalized in each case so that the initial energy released to the heaters matches experiment (see Methods). Inertial simulations for nanodots are shown just in a short time window due to their high computational cost. Also shown are the experimental and theoretical changes in diffraction efficiency for close-packed, i.e. $(P - L) < 2\ell$, **(c)** nanolines (1D) and **(d)** nanodots (2D) of different sizes L and periods P. Theoretical results are solutions of Eq. (1) with $\ell_{eff} = (P - L)/2$ while the other parameters are the same size-independent values. The only fitting parameter for the entire data set is the intrinsic thermal boundary resistance, which is set to 2.25 $nKm^2/W$ for this work. The excellent agreement between KCM and the experimental data for the highly non-diffusive decay for both 1D- and 2D-confined heat source geometries demonstrates the predictive capability of this model.*

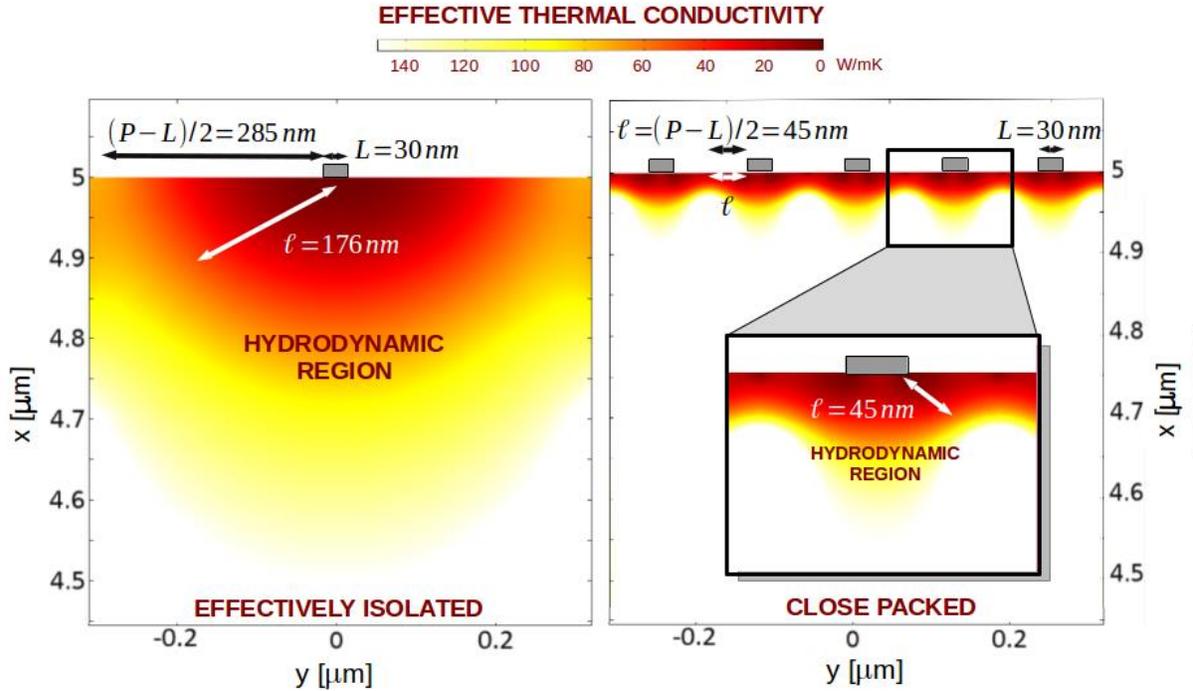

*Figure 3. Hydrodynamic regions in effectively isolated and close-packed situations. Effective thermal conductivity profile on silicon, $|q|/|\nabla T|$, predicted by KCM for a nanoheater of width 30nm at t=0.5ns for **(left)** isolated (P=600nm) and **(right)** close-packed (P=120nm) configurations. Similar to fluids, a friction-like reduction of thermal transport appears in the regions of the substrate where heat flux gradients are large. Parameter $\ell$ defines the characteristic size of the region below heaters where these hydrodynamic effects are important (hydrodynamic region). When sources are separated a distance larger than $2\ell$ (effectively isolated lines) one uses the intrinsic value $\ell=176nm$. When this distance is smaller, i.e. $(P - L) < 2\ell$, an effective value $\ell_{eff} = (P - L)/2\ (< \ell)$ is used. The red color indicates regions where the thermal transport has been reduced (compared to diffusion) while the white color represents regions of diffusive transport. In close-packed configurations, the interaction between heaters homogenizes the profile thus reducing viscous effects to a smaller region of size $\ell_{eff}$. As a result, close-packed configurations evacuate heat faster than isolated lines of the same width as shown in Ref.[8]. The profiles shown do not appreciably change during the timescale of experiments. Note that scales are the same in both panels.*

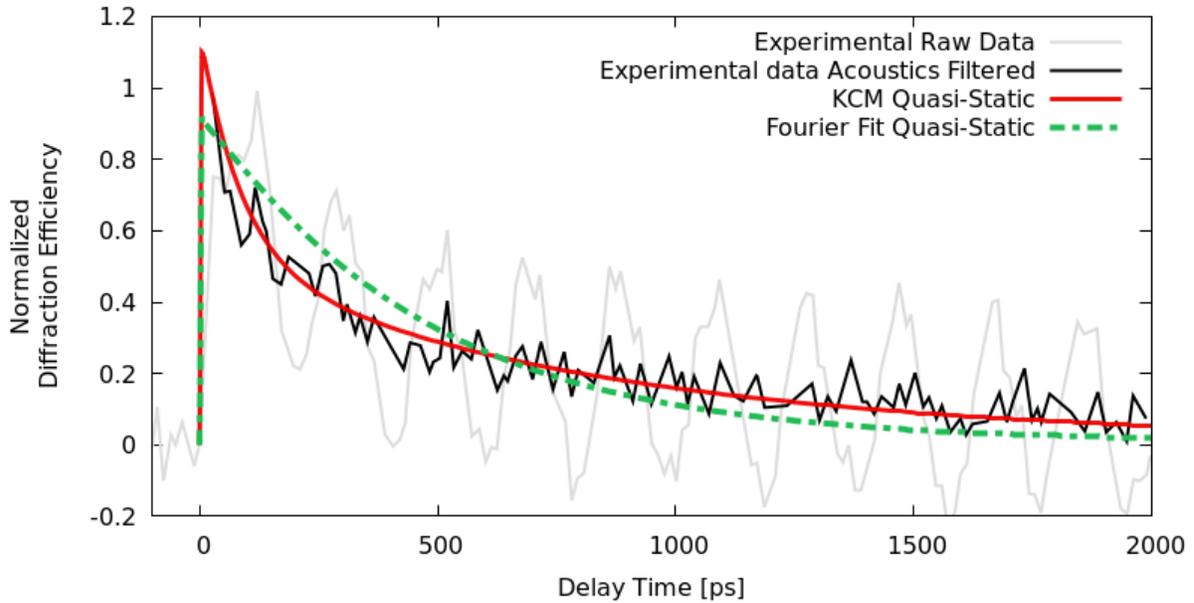

*Figure 4. Experimental and theoretical quasi-static change in diffraction efficiency. Comparison of the thermal relaxation for effectively isolated heater lines of L=250 nm and P=1000 nm. The black (grey) line denotes experimental data without (with) acoustics, the red line is our new KCM prediction using intrinsic parameters, while the green line is a Fourier model using an effective thermal boundary resistance value fitted to obtain the best match to data. The Fourier fit overestimates experimental decay at short times and underestimates it at long times. Experimental measurements indicate that the thermal decay of heaters cannot be described by just one characteristic time, like the prediction by Fourier's model; however, KCM captures the decay for all times. Raw experimental data is from Ref.[8].*

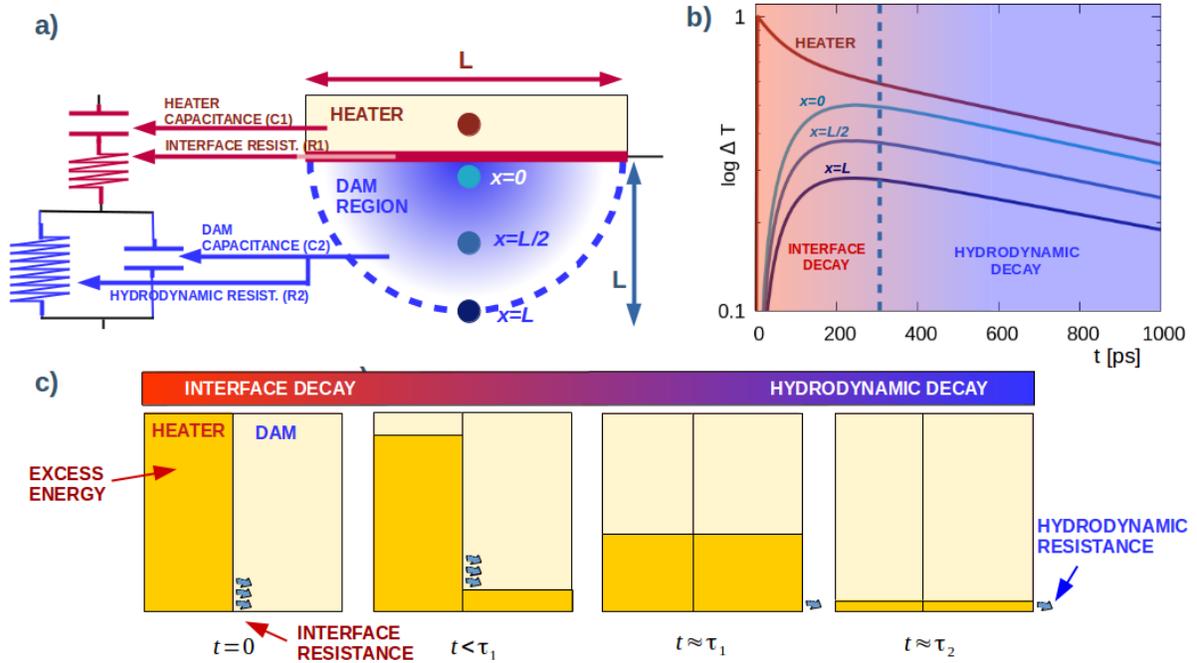

*Figure 5. Two-box model for the thermal decay of heaters for $L < \ell$. (a)* The energy released by the heater (with heat capacity per unit surface, $C_1$) crosses the interface with the substrate at a rate determined by the thermal boundary resistance, $R_1$. The thermal response of the substrate is determined by a region of size L below the heater—the dam region—which acts both as a heat reservoir of capacity $C_2$, and as a thermal resistance $R_2$ due to viscosity from hydrodynamic effects. An analogy to an equivalent electrical circuit is shown. *(b)* The temperature as a function of time is shown for the positions indicated in (a) from KCM solutions for $L = 30$nm and $P = 600$nm. At short times, the dam region retains the energy released by the heater and increases in temperature with a time scale $\tau_1$ dominated by the interface resistance. At larger times, a slow joint decay of heater and dam temperatures occurs with a characteristic time $\tau_2$ determined by the hydrodynamic resistance $R_2$. *(c)* Cartoon of the two-box model in analogy with fluids. The two boxes represent the heater and dam with the water level indicating the temperature. For times less than $\tau_1$, the excess energy flows out of the heater into the dam through the interface resistance until temperatures equilibrate. For times on scale of $\tau_2$, excess energy in heater and dam escapes to the rest of the substrate at a rate ruled by hydrodynamic effects.

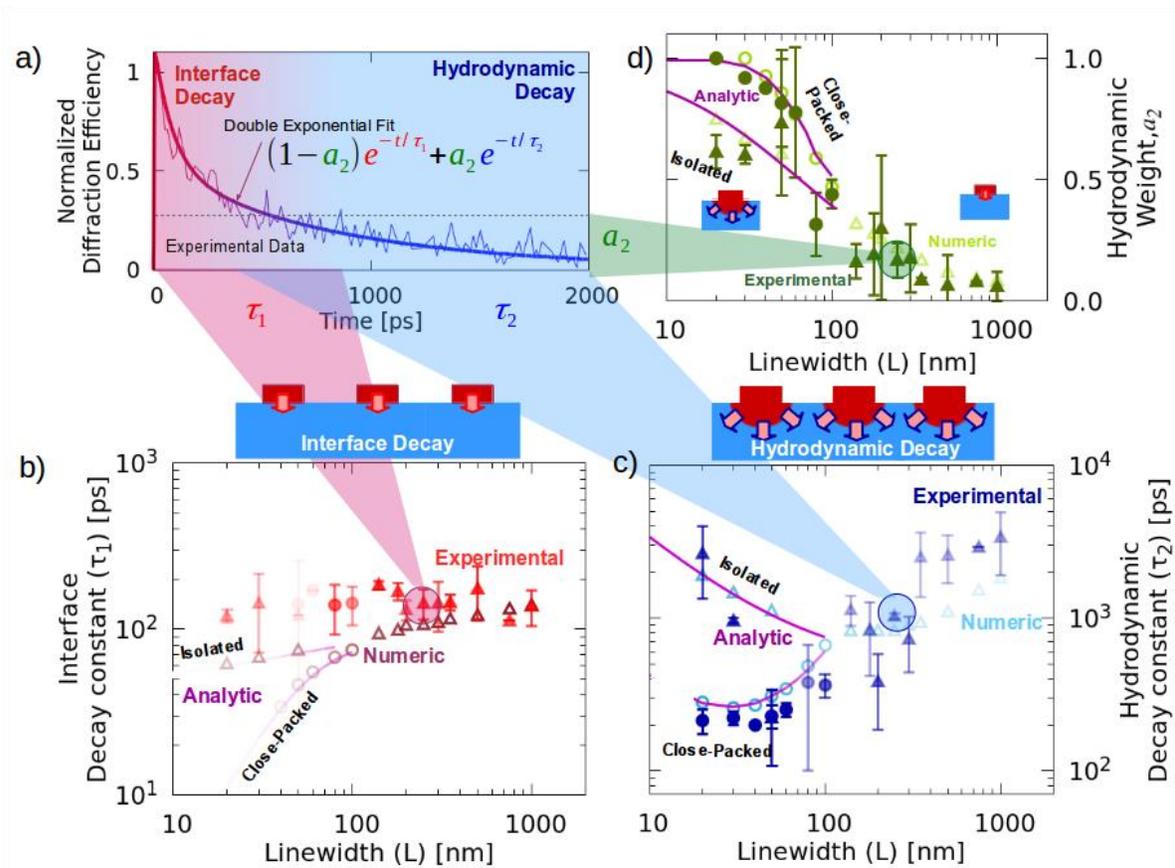

*Figure 6. Two characteristic decay times in thermal relaxation of nanoline (1D) experiments. (a) The experimental change in diffraction efficiency, with oscillations removed, for a heater line of L=250 nm and P=1000 nm (thin line) can be fitted with a double exponential decay (thick line), from which two characteristic times are extracted: a short time scale (red line region, $\tau_1$) and a long time scale (blue line region, $\tau_2$). (b,c) Characteristic time $\tau_1$ and $\tau_2$ versus heater linewidths L for effectively isolated (triangles) and close-packed (circles) experiments. KCM numerical (analytical) results are denoted by open symbols (lines). The color intensity in the symbols indicates the weight of each characteristic time in the overall decay. The short time scale ($\tau_1 \sim 0.1 ns$) is dominated by the interface resistance, while the long one ($\tau_2 \sim 1 ns$) is ruled by the hydrodynamic effects in the substrate. Additionally, the difference between the dissipation of close-packed versus effectively isolated heat sources is demonstrated. (d) The normalized weight of the hydrodynamic characteristic time in the temperature decay, $a_2$ ($= 1 - a_1$), is displayed versus linewidth for all experiments, showing the transition from interface- to hydrodynamic-*

*dominated decay as source size decreases. These experimental fits include raw data from Refs.[8,9] and the current study.*

# Supplementary Information

Thermal transport from 1D- and 2D-confined nanostructures on silicon probed using coherent extreme UV light: General and predictive model yields new understanding

# Contents

# 1    Thermoelastic Model

The model equations can be solved using Finite Elements methods to determine the evolution of the displacement vector $\vec{u}$, the temperature $T$ and the heat flux $\vec{q}$ in the metalic domains and the substrate, which allow comparision with the experimental measurements. The set of equations is the following:

### Linear Elastic Equations

To describe the mechanic evolution of the system we use the elastic equation including inertial effects [15]

$$\rho \frac{\partial^2 \vec{u}}{\partial t^2} = \nabla \cdot \sigma, \tag{1}$$

where $\rho$ is the density and $\sigma$ is the stress tensor of the material.

For the heaters we use the bulk nickel stress tensor (isotropic) with linear thermal expansion:

$$\sigma_{ik} = K_{\text{Ni}}[-\alpha_{\text{Ni}}(T - T_0) + \nabla \cdot \vec{u}]\delta_{ik} + \mu_{\text{Ni}}\left[\frac{\partial u_i}{\partial x_k} + \frac{\partial u_k}{\partial x_i} - \frac{2}{3}\nabla \cdot \vec{u}\,\delta_{ik}\right] \quad (2)$$

where $\alpha_{\text{Ni}}$, $K_{\text{Ni}}$ and $\mu_{\text{Ni}}$ are the nickel coefficient of thermal expansion, the compressibility modulus and the shear modulus, respectively, and $T_0 = 300K$ is the ambient temperature.

For the silicon stress tensor, we also assume linear thermal expansion and we use an anisotropic form to include the effects coming from the stress generated by the nanolines on the substrate top surface [20]. The characterization of this stress tensor has been reported elsewhere [16] and it is necessary to reproduce the frequency of the mechanic oscillations:

$$\sigma = D : [\nabla \vec{u} - \alpha_{\text{Si}}(T - T_0)\mathbb{I}] \quad (3)$$

where $\alpha_{\text{Si}}$ is the silicon coefficient of thermal expansion, $\mathbb{I}$ is the identity matrix and $D$ is the anisotropic elasticity matrix.

In order to obtain the quasi-static relaxation (i.e. suppress the acoustic oscillations) we substitute equation (1) by $0 = \nabla \cdot \sigma$.

## Heat Transport Equations

Heat transport is modeled including the corrections of the phonon hydrodynamic model in the region where heat is carried mainly by phonons i.e. the silicon substrate.

**Nickel Nanostructures (Fourier).** Heat transport in metal lines or dots (height $h = 11.5nm$ and width $L$) is carried by electrons. As their mean free paths are much shorter than the sizes of the nanostructures, nonlocal effects are not expected. Consequently, Fourier's law is valid on these domains, in which we denote the heat flux and the temperature with subindex 1.

We include the linear thermoelastic coupling in the energy conservation equation to model the transfer between thermal and elastic energy:

$$c_{\text{Ni}}\frac{dT_1}{dt} + \nabla \cdot \vec{q_1} = Q - p_{\text{Ni}}\frac{\partial}{\partial t}\nabla \cdot \vec{u} \quad (4)$$

where $c_{\text{Ni}}$ is the specific heat of Nickel and $p_{\text{Ni}} = \alpha_{\text{Ni}} K_{\text{Ni}} T_0$. The heating energy density $Q$ is $1W/m^3$ for time $t < 2.5ps$ and 0 otherwise. The normalized thermal response of the system does not depend on the used value for $Q$ due to the linearity of the model equations. The temporal duration of the pulse is estimated using a two-temperature model [14, 2]. Moreover, we confirmed that the exact length of the heat pulse (0-10ps duration) does not significantly affect the results.

For the transport equation we use the Fourier's law

$$\vec{q}_1 + \kappa_{\text{Ni}} \nabla T_1 = 0 \tag{5}$$

where $\kappa_{\text{Ni}}$ is the bulk thermal conductivity of Nickel.

**Silicon substrate (Hydrodynamic Transport).** We use the energy conservation equation with linear thermoelastic coupling and without the source term,

$$c_{\text{Si}} \frac{dT}{dt} + \nabla \cdot \vec{q} = -p_{\text{Si}} \frac{\partial}{\partial t} \nabla \cdot \vec{u} \tag{6}$$

where $c_{\text{Si}}$ is the specific heat of silicon and $p_{\text{Si}} = \alpha_{\text{Si}} K_{\text{Si}} T_0$.

For the heat transport we use the GKE including non-local and memory effects,

$$\tau \frac{\partial \vec{q}}{\partial t} + \vec{q} + \kappa_{\text{Si}} \nabla T = \ell^2 (\nabla^2 \vec{q} + \alpha \nabla \nabla \cdot \vec{q}) \tag{7}$$

where $\kappa_{\text{Si}}$ is the silicon bulk thermal conductivity, $\ell$ is the non local length, $\tau$ is the heat flux relaxation time and $\alpha$ is a dimensionless parameter. Equation (7) can be derived from the BTE assuming an averaged phonon mode relaxation time with $\alpha = 1/3$ [7] and for the collective regime (normal dominant phonon collisions) with $\alpha = 2$ [8]. In consistency with [21, 18, 3, 4], here we propose the use of an effective form of equation (7) by considering interpolated parameters $\kappa, \ell, \tau$ between the kinetic (resistive dominant phonon collisions) and the collective situations at the reference temperature $T_0$, and $\alpha = 1/3$. Details of the interpolation and the ab initio calculation of these parameters can be found elsewhere [19, 18].

## Boundary Conditions

**Nanostructure Free surfaces** Thermal insulation is imposed by setting to zero the normal heat flux component

$$\vec{q}_1 \cdot \vec{n} = 0 \tag{8}$$

where $\vec{n}$ is the boundary normal vector.

**Interface** On one hand we impose continuity of the normal component of the heat flux going through the interface:

$$\vec{q}_1 \cdot \vec{n} = \vec{q} \cdot \vec{n} \tag{9}$$

where $\vec{n}$ is the interface normal vector pointing towards the semiconductor.

On the other hand, we impose the temperature jump boundary condition that accounts for the nonequilibrium effects introduced by the interface (assuming diffusive reflections):

$$(T - T_1) = -R_1 \vec{q} \cdot \vec{n} + \frac{1}{\gamma^{\text{Si}}} (\beta \nabla \cdot \vec{q} - \nabla \vec{q} : \chi) \tag{10}$$

where $\beta, \chi$ are ab initio calculated characteristic lengths. Derivation of this boundary condition and the explicit microscopic expression for these parameters by imposing energy balance restrictions with the use of the specific non-equilibrium distribution function of each domain can be found elsewhere [4]. The tensor $\chi$ is diagonal. A lower bound for the thermal resistance value $R_1$ is also obtained in [4] by assuming phonon diffusive boundary scattering and perfect interface (no reduction of the contact area between nickel and silicon due to fabrication defects),

$$R_1^{\text{min}} = \frac{1}{2}(\frac{1}{\gamma_0^{\text{Si}}} + \frac{1}{\gamma_0^{\text{Ni}}}). \tag{11}$$

where $\gamma_0^{\text{Si}} = \frac{1}{4}c_{\text{Si}}v_{\text{Si}}$ and $\gamma_0^{\text{Ni}} = \frac{1}{4}c_{\text{Ni}}v_{\text{Ni}}$ (with $v$ the average phonon group velocity for each material respectively). The actual thermal resistance value $R_1$ is unknown due to the lack of knowledge of the interface defects. Therefore we fit a correction factor $R_1/R_1^{\text{min}} = 3.11$ to the data obtained from the largest experimentally available gratings $L > 750$nm, in which hydrodynamic effects are not relevant (i.e. Fourier and KCM predictions coincide). Hence, this correction does not depend on the model used. The second and third terms in the right-hand side of expression (10) are required to properly describe the non-local effects induced by the interface in the substrate heat flux and temperature profiles. In the present work only cause minor corrections for the smallest linewidths $L < 50$nm. The same correction factor has been applied to the resistance weighting the non-local term $\gamma_0^{\text{Si}}/\gamma^{\text{Si}} = 3.11$.

**Slip boundary conditions** For silicon, where a second order derivative of the heat flux is included on the thermal transport equation, a boundary condition for the tangential flux $\vec{q}_t$ is required. We use the slip boundary condition both in the silicon interface and free surfaces [1, 3]

$$\vec{q}_t = C\ell \, \nabla \vec{q}_t \cdot \vec{n} \tag{12}$$

where

$$C = \frac{1+p}{1-p}, \tag{13}$$

$p$ is the fraction of specularly reflected phonons in the boundaries (details can be found in [3]). In consistency with (10) here we consider diffusive reflections $p = 0$, i.e. $C = 1$. In the silicon free surface, insulation is also imposed

$$\vec{q} \cdot \vec{n} = 0 \tag{14}$$

**Substrate** Only a geometry periodically repeated unit is simulated by imposing periodic boundary conditions. The reference temperature $T_0 = 300$K and null displacement vector $\vec{u} = 0$ is fixed in the substrate base.

Table 1: Thermoelastic model parameter values at 300K.

|  | Nickel | Silicon |
|---|---|---|
| $\rho$ [Kg m$^{-3}$] | 8900 | 2329 |
| $\alpha$ [K$^{-1}$] | $12.77 \cdot 10^{-6}$ | $3 \cdot 10^{-6}$ |
| $K$ [Pa] | $175 \cdot 10^9$ | $95 \cdot 10^9$ |
| $\mu$ [Pa] | $76 \cdot 10^9$ | $52 \cdot 10^9$ |
| $c$ [J m$^{-3}$ K$^{-1}$] | $4 \cdot 10^6$ | $1.6 \cdot 10^6$ |
| $\kappa$ [W m$^{-1}$ K$^{-1}$] | 91 | 145 |
| $\tau$ [s] | - | $50 \cdot 10^{-12}$ |
| $\ell$ [m] | - | $176 \cdot 10^{-9}$ |

## Parameter Values

Regarding the interface boundary condition, the calculated conductance bounds are $\gamma_0^{\text{Si}} = 1068\text{MW}/\text{m}^2\text{K}$ and $\gamma_0^{\text{Ni}} = 1960\text{MW}/\text{m}^2\text{K}$, and the thermal boundary resistance used is $R_1 = 2.25\text{nK}\,\text{m}^2/\text{W}$. The tensor $\chi$ is diagonal with $\chi_{xx} = -31\text{nm}$, $\chi_{yy} = \chi_{zz} = -16\text{nm}$ where $x$ denote the normal direction pointing towards the semiconductor. The length $\beta = -21\text{nm}$.

Regarding the silicon stress tensor, we use an anisotropic model in Voigt notation $\{yy, xx, zz, xz, yz, zx\}$ with the following rotated tensor $D_{11} = 203, D_{12} = 66.5, D_{22} = 173, D_{13} = 36.5, D_{23} = 66.5, D_{33} = 203, D_{14} = 0, D_{24} = 0, D_{34} = 0, D_{44} = 83, D_{15} = 0, D_{25} = 0, D_{35} = 0, D_{45} = 0, D_{55} = 53, D_{16} = 0, D_{26} = 0, D_{36} = 0, D_{46} = 0, D_{56} = 0, D_{66} = 83$ [GPa] to match the simulated geometry that $y$ is along [001] and $x$ (interface normal direction) is along [110] in the silicon crystal.

The rest of parameter values used in the model for each material can be found in the Table 1.

## 2 Thermal Decay Analysis

In this section, we compare the non-equilibrium evolution of the system according to the Fourier model and to the Kinetic Collective Model (KCM) and we derive the two-box model equations from the KCM analytical solutions in the Stokes regime ($L < \ell$).

## Fourier Model

First consider diffusive heat transport both in the Si substrate and in the Ni heater along with a Kapitza interface boundary condition with resistance $R$. We use the bulk thermal properties from Table 1 and we denote the thermal diffusivity of Nickel $\chi_{\text{Ni}} = \kappa_{\text{Ni}}/c_{\text{Ni}} = 2.2 \cdot 10^{-5}\text{m}^2\,\text{s}^{-1}$ and of silicon $\chi_{\text{Si}} = \kappa_{\text{Si}}/c_{\text{Si}} = 9 \cdot 10^{-5}\text{m}^2\,\text{s}^{-1}$. For illustration purposes, we discuss this benchmark model considering the spe-cific case of heater height $h = 10$nm and width $L = 20$nm, with $R = 1$nK m$^2$/W. In Figure 1 we show the corresponding temperature evolution of the heater obtained with COMSOL Multiphysics.

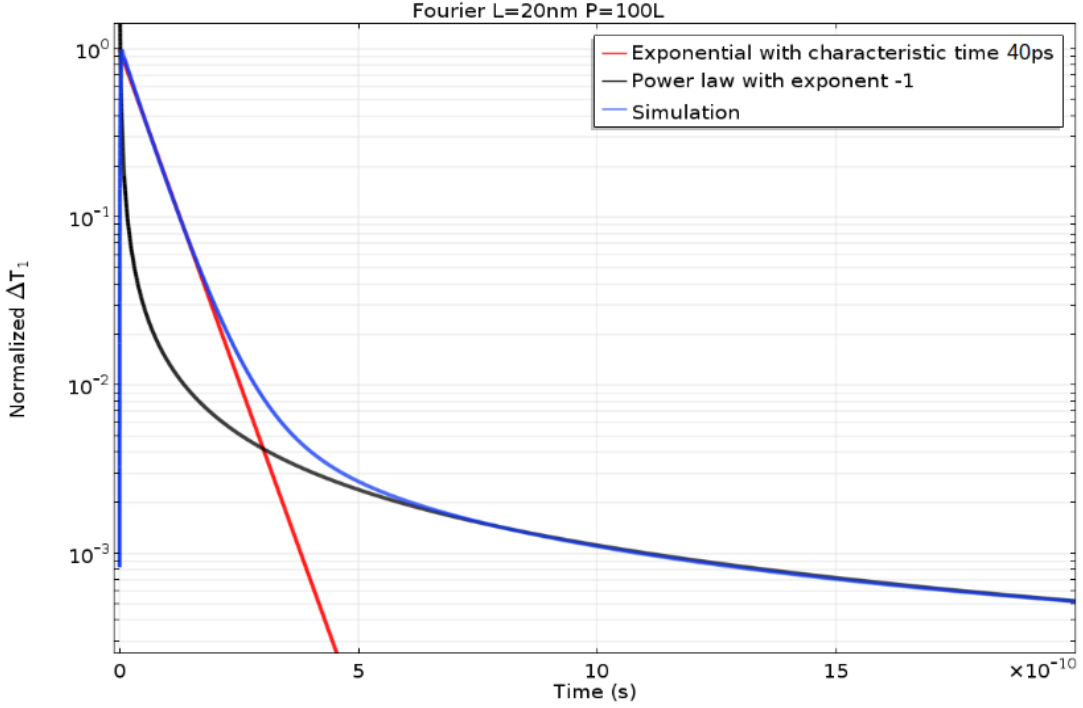

**Figure 1:** Heater temperature evolution for $h = 10$nm, $R = 1$nK m$^2$/W, $L = 20$nm and $P = 100L$ according to Fourier model. The initial decay is fitted using an exponential and the long-time decay is fitted using a power-law.

The time scale of the thermal evolution in the heater is extremely fast $h^2/\chi_{\text{Ni}} = 4.5$ps and hence the temperature in the heater is almost uniform within the time scale of the experiment. In the substrate, at time $t$, diffusion has penetrated a region of size $\sqrt{t\chi_{\text{Si}}}$. We can quantify an effective thermal resistance due to diffusion $r(t) = \sqrt{t\chi_{\text{Si}}}/\kappa_{\text{Si}}$. At early times $R > r(t)$, so the thermal decay is dominated by the interface. At times larger than $R^2\kappa_{\text{Si}}c_{\text{Si}} = 232$ps, we have $r(t) > R$ thus the thermal decay is dominated by substrate diffusion.

For $t < R^2\kappa_{\text{Si}}c_{\text{Si}}$: The heat flux in the interface is $|\vec{q}| = \Delta T/R$, where $\Delta T$ is the temperature difference between the heater and the substrate across the interface. Moreover, the heat flux leaving the heater can be estimated as $|\vec{q}| \sim -c_{\text{Ni}}h\frac{dT}{dt}$, and hence $\Delta T \sim -c_{\text{Ni}}hR\frac{dT}{dt}$. Therefore, the temperature evolution of the heater is an exponential with characteristic time $\tau_F = c_{\text{Ni}}hR = 40$ps.

For $t > R^2\kappa_{\text{Si}}c_{\text{Si}}$: Substrate diffusion has no characteristic time scale (infinite substrate) and hence the thermal decay follows a power law with an exponent depending on the space dimensionality. The thermal evolution in the region below the heater is instantaneous $L^2/\chi_{\text{Si}} = 4.4$ps and hence we can consider that the heated region is a point and the temperature evolution is 2D (exponent -1).

In summary, the Fourier model predicts an initial exponential thermal decay with characteristic time $\tau_F = c_{Ni}hR = 40$ps followed by a power law decay with exponent -1. The transition between both decays is estimated to be at time $R^2\kappa_{Si}c_{Si} = 232$ps.

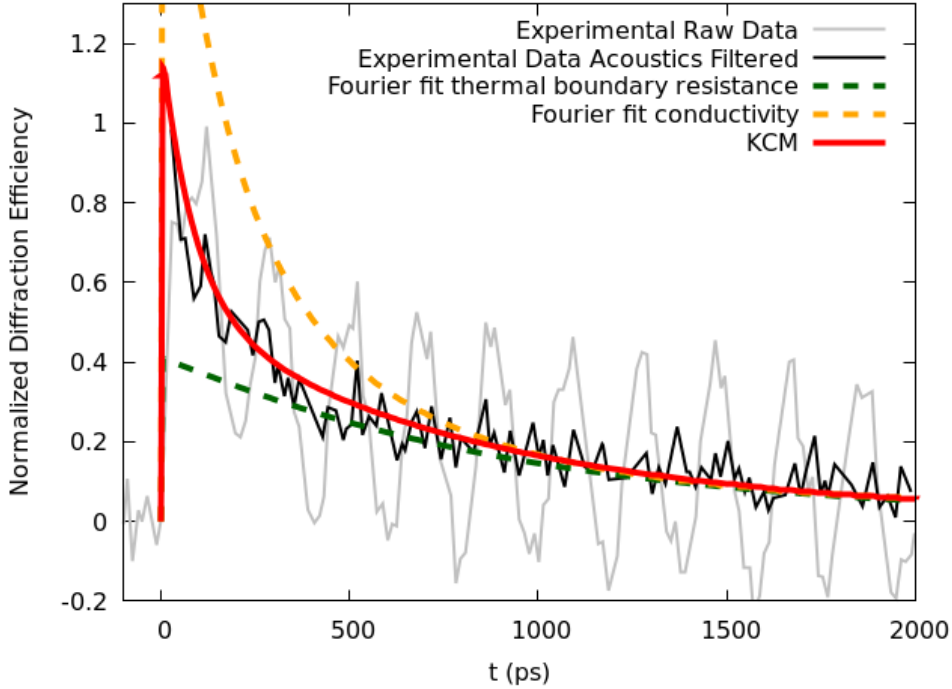

**Figure 2:** Fourier fits to experimental data restricted to $t > 500ps$ for a heater line of 250nm. The TBR fit ($R = 19$nK m$^2$/W) using the intrinsic value for the substrate conductivity underpredicts the temperature for $t < 500ps$, whereas the conductivity fit ($\kappa = 0.4\kappa_{bulk}$) using the intrinsic value for the TBR overpredicts it. In both models, the predicted fraction of energy evacuated from the heater is substantially distorted.

As shown in Figure 4 of the main text, the functional form of the thermal decay according to Fourier's law is not consistent with the experimental decay from EUV scatterometry measurements, where a double exponential decay with two distinct characteristic times is observed (see also Figure 6 of the main text). Care must be taken when comparing our results with those from time-domain thermal reflectance (TDTR) [12], which are two completely different techniques. One key difference is that visible-based probe experiments often need to limit the analysis to the region $t > 500ps$ due to the challenges separating the contributions of out-of-equilibrium electrons and thermal decay. However, our EUV probe does not suffer from this limitation as we measure the surface deformation through diffraction, which is not altered by the presence of nonequilibrium electrons. If we take EUV scatterometry data and exclude from our analysis the measurements for $t < 500ps$, one of the decay times (∼100ps) reported in this work cannot be observed. Consequently, a diffusive model could fit the experimental data by excluding the initial timescales, as done in previous works like [12]. In Figure 2, we show two different fits using Fourier with an effective TBR and an effective substrate conductivity, respectively, restricted to $t > 500ps$. Both approaches can reproduce the tail of the decay but fail to reproduce the initial system response. After the first nanosecond, these models are able to reproduce just the last 20% of the signal amplitude but cannot predict the other 80%, i.e. the largest part of the energy dissipation from heaters which is also the most important for applications.

## Kinetic Collective Model

We assume now diffusive heat transport in the heater and hydrodynamic heat transport in the substrate using the Kinetic Collective Model as explained in Supplementary Section 1. In figure 3, we show an example of the temperature evolution of the heater for $L = 30$nm and $P = 400$nm obtanied with COMSOL.

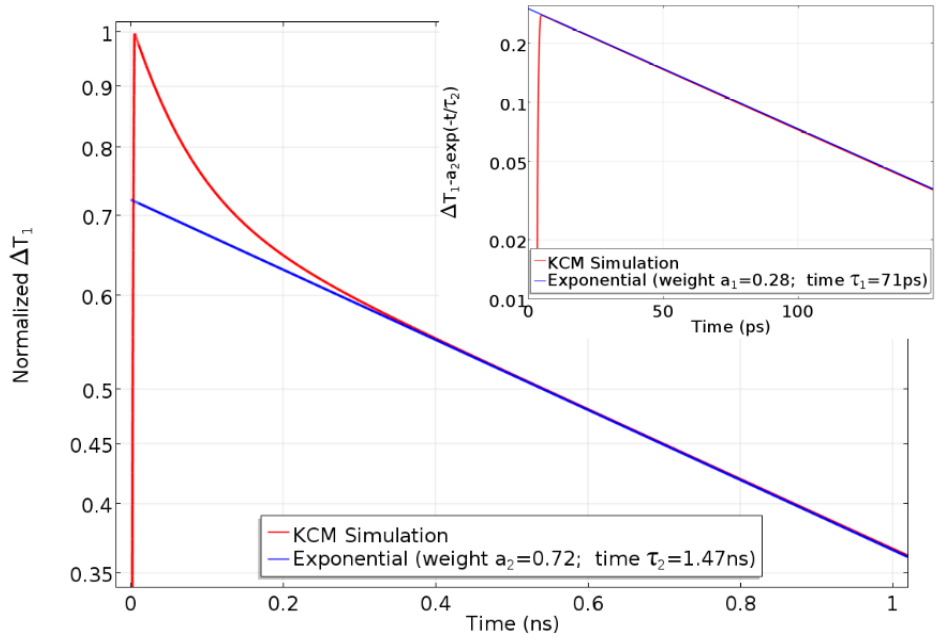

**Figure 3:** Heater temperature evolution for $L = 30$nm and $P = 400$nm according to KCM. The thermal decay is fitted with a double exponential.

In contrast to the Fourier model, the KCM predicts a slower thermal decay that can be fitted using a double exponential within the time scale of the experiment. We denote as $\tau_1, \tau_2, a_1, a_2$ the characteristic times and weights of the first and the second exponentials, respectively. The geometrical dependencies of these coefficients are represented in Figure 6 of the main text.

**Description of Figure 6 and sensitivity of the KCM parameters**

The information condensed in Figure 6 of the main article (reproduced here as Figure 4) is more easily understandable by identifying all the double exponential decay coefficients for an specific sample and analyzing the sensitivity of the decay to changes in the KCM parameters values. Here we consider two examples that represent two different situations. The selected cases are represented in blue and red in Figure 4.

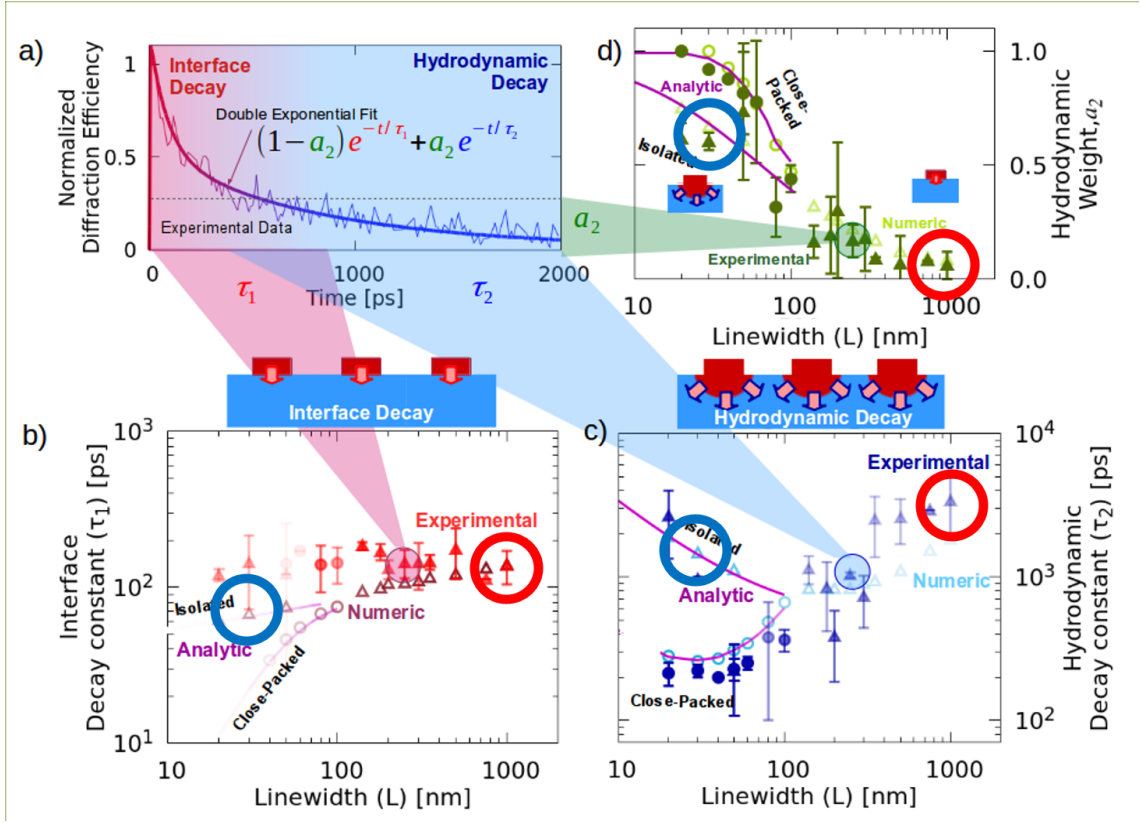

**Figure 4:** Two characteristic decay times in thermal relaxation of nanoline (1D) experiments. (a) The experimental change in diffraction efficiency, with oscillations removed, for a heater line of $L = 250$nm and $P = 1000$nm (thin line) can be fitted with a double exponential decay (thick line), from which two characteristic times are extracted: a short time scale (red line region, $\tau_1$) and a long time scale (blue line region, $\tau_2$). (b,c) Characteristic time $\tau_1$ and $\tau_2$ versus heater linewidths $L$ for effectively isolated (triangles) and close-packed (circles) experiments. KCM numerical (analytical) results are denoted by open symbols (lines). The color intensity in the symbols indicates the weight of each characteristic time in the overall decay. The short time scale ($\tau_1 \sim 0.1$ns) is dominated by the interface resistance, while the long one ($\tau_2 \sim 1$ns) is ruled by the hydrodynamic effects in the substrate. Additionally, the difference between the dissipation of close-packed versus effectively isolated heat sources is demonstrated. (d) The normalized weight of the hydrodynamic characteristic time in the temperature decay, $a_2(=1-a_1)$, is displayed versus linewidth for all experiments, showing the transition from interface- to hydrodynamic-dominated decay as source size decreases.

The blue circles indicate the coefficients of the double exponential decay for isolated small heaters with size $L = 30$nm and periodicity $P = 400$nm ($\tau_1 = 68$ps, $\tau_2 = 1470$ps, $a_2 = 1 - a_1 = 0.7$). The thermal evolution for this sample is repre-sented with blue lines in the two top plots of Figure 5. The influence of modifying the non-local length $\ell$ or the boundary resistance value $R_1$ is displayed in the left and right plots, respectively. Note that the smaller time $\tau_1$ is determined by $R_1$ and the larger one $\tau_2$ by $\ell$, so each decay is associated to a different mechanism. In this case the weight $a_2 > 0.5$ and the hydrodynamic time is larger than the TBR dominated time ($\tau_2 > \tau_1$). Therefore, the heater thermal evolution is more sensitive to the $\ell$ value than to the interfacial resistance value $R_1$ as shown in Figure 5.

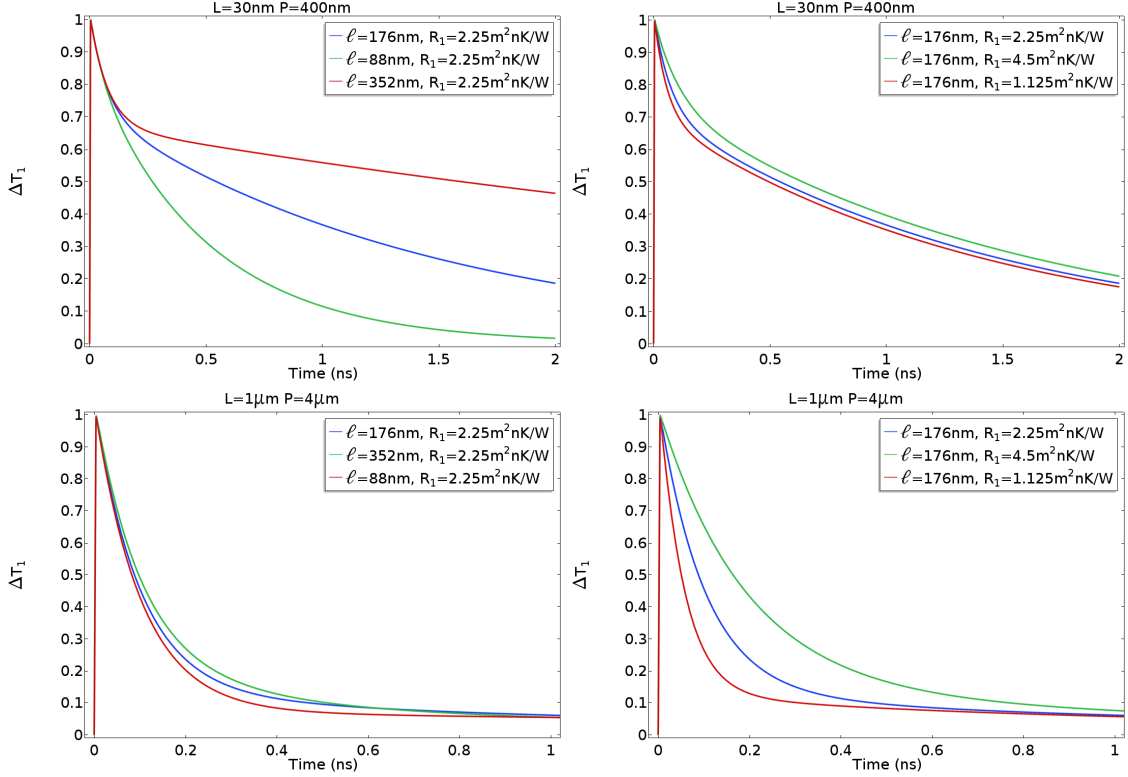

**Figure 5:** Sensitivity of KCM parameters. Left Plots: Thermal decay using the values given by our KCM-ab initio model (blue line) in comparison with the same system using a non-local length multiplied by two (green) and divided by two (red) times. Right Plots: Analogous comparison is provided but changing the used value of the resistance $R_1$. Top: Small heater size ($L = 30$nm and $P = 400$nm). Bottom: Large heater size ($L = 1\mu$m and $P = 4\mu$m). Notice that the boundary resistance controls the initial decay $\tau_1$ and the non-Fourier conduction controls the decay at larger times $\tau_2$. Moreover, the weight of the second exponential with time $\tau_2$ increases by decreasing the line-width.

The red circles in Figure 4 identify the coefficients for large heaters with size $L = 1\mu$m ($\tau_1 = 139$ps, $\tau_2 = 1840$ps, $a_2 = 0.09$). In this sample the situation is the contrary with respect to the previous case. The hydrodynamic weight is small ($a_2 < 0.1$) and the timescale $\tau_1$ determined by the TBR dominates. In the bottom plots of Figure 5 the decay for this sample is represented in blue, along with the same sensitivity analysis of the KCM parameters. Notice that the decay is mainly influenced by $R_1$ while the change of $\ell$ does not have much effect. This is expected since for large heater sizes the decay is well described by Fourier's law along a Kapitza interfacial resistance.

Therefore, there is a transition from the thermal relaxation dominated by hydrodynamic effects observed for small heater sizes, to the evolution dominated by the interfacial resistance for large sizes. For intermediate sizes, the two mechanisms are important and a double exponential decay is evident (see Figure 4 of the main text). The experimental validation of this transition is displayed in Figure 6d of the main text.

## The Two-Box Model: Analytical Derivation of double exponential thermal decay.

Here we derive the analytical expression for the parameters of the double exponential thermal decay predicted by KCM in the case $L < \ell$. We denote $x$ the cross-plane direction towards the substrate and $y$ the in-plane direction. The origin of coordinates is the center of the interface.

In KCM, the heat flux in the substrate is described through the hydrodynamic heat transport equation

$$-\kappa_{\text{Si}} \nabla T = \tau \frac{\partial \vec{q}}{\partial t} + \vec{q} - \ell^2 (\nabla^2 \vec{q} + \alpha \nabla (\nabla \cdot \vec{q})) \tag{15}$$

along with the energy conservation equation

$$\nabla \cdot \vec{q} = -c_{\text{Si}} \frac{\partial T}{\partial t}. \tag{16}$$

We neglect the term $\tau \frac{\partial \vec{q}}{\partial t}$ because it does not play a significant role in the present experimental conditions. We also neglect the thermo-elastic coupling in the energy conservation for simplicity. We consider the Stokes regime $L < \ell$ (heat transport dominated by viscosity); then one can neglect the term $\vec{q}$ in (15) close to the heater since we expect $\ell^2 \nabla^2 \vec{q} \sim \frac{\ell^2}{L^2} \vec{q}$. Is worth to note that the heat flux profile saturates after a fast transient and viscosity remains constant during the rest of the experiment. The heat flux profile saturation time can be estimated as $\ell^2/\chi_{\text{Si}} = 341 ps$. To illustrate this, in Figure 6 we show the time evolution of the different terms in equation (15) at $x = L$ and $y = 0$ with $L = 20$nm.

We have
$$\frac{\kappa_{\text{Si}}}{\ell^2} \nabla T = \nabla^2 \vec{q} + \alpha \nabla (\nabla \cdot \vec{q}). \tag{17}$$

Now we perform integration in the region dominated by viscous effects $x < 2\ell$:

$$\frac{\kappa_{\text{Si}}}{\ell^2} \int_0^{2\ell} \nabla T dx = -\frac{\kappa_{\text{Si}}}{\ell^2}(T_2 - T_2^{2\ell}) = \int_0^{2\ell} dx \left(\frac{\partial^2 q_x}{\partial x^2} + \frac{\partial^2 q_x}{\partial y^2}\right) + \alpha \int_0^{2\ell} dx \frac{\partial \nabla \cdot \vec{q}}{\partial x}, \tag{18}$$

where $T_2$ is the substrate temperature at the interface $x = 0$ and $T_2^{2\ell}$ is the substrate temperature at $x = 2\ell$.

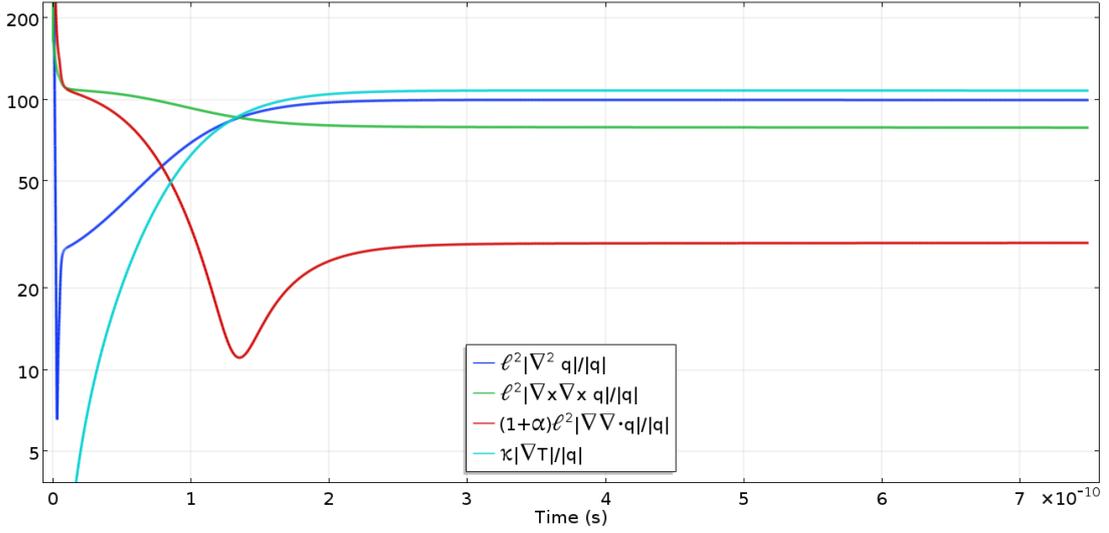

**Figure 6:** Time evolution of the hydrodynamic heat transport equation terms at $x = L$ and $y = 0$ with $L = 20$nm.

At the time scales considered in the experiment $T_2^{2\ell}$ is constant and close to the initial temperature $T_2^\infty = T_0$. Moreover, during the experimental time scale, the heat flux and its derivatives at $x = 2\ell$ are neglectable in front of the heat flux and its derivatives at the interface $x = 0$.

Therefore, we can perform integration of (16) in the hydrodynamic region to obtain
$$\int_0^{2\ell} dx \frac{\partial \nabla \cdot \vec{q}}{\partial x} = c_{\text{Si}} \frac{\partial T_2}{\partial t}, \tag{19}$$

or, equivalently,
$$\int_0^{2\ell} dx \frac{\partial^2 q_x}{\partial x^2} = c_{\text{Si}} \frac{\partial T_2}{\partial t} + \frac{\partial q_y}{\partial y}\bigg|_{x=0}. \tag{20}$$

Introducing (19,20) in (18), we obtain
$$\frac{\kappa_{\text{Si}}}{\ell^2}(T_2 - T_2^\infty) + (1+\alpha)c_{\text{Si}} \frac{\partial T_2}{\partial t} = -\int_0^{2\ell} dx \frac{\partial^2 q_x}{\partial y^2} - \frac{\partial q_y}{\partial y}\bigg|_{x=0}. \tag{21}$$

Now we average equation (21) over all the interface points $y \in [-L/2, L/2]$.
$$\frac{\kappa_{\text{Si}}}{\ell^2}(\bar{T}_2 - T_2^\infty) + (1+\alpha)c_{\text{Si}} \frac{\partial \bar{T}_2}{\partial t} = \frac{B}{L} \bar{q}_x \tag{22}$$

where $\bar{T}_2$, $\bar{q}_x$ are the average temperature and heat flux in the interface, respectively, and
$$B = -\frac{\int_0^{2\ell} dx \int_{-L/2}^{L/2} dy \frac{\partial^2 q_x}{\partial y^2}}{\bar{q}_x} - \frac{2 q_y\big|_{y=L/2 \,\&\, x=0}}{\bar{q}_x} \tag{23}$$

is a geometric parameter related with the average heat flux profile in the hydrodynamic region. This parameter saturates because the heat flux profile reach a stationary situation as can be seen in Figure 3. The saturated value can be estimated from the COMSOL simulations. We obtain a constant value for $L < \ell$ in the 1D geometry: $B = 3$.

Diffusion in the heater region is extremely fast and hence the temperature of the heater $T_1$ is uniform within the time scale of the experiment. Therefore, using the averaged form of a Kapitza interface boundary condition $\bar{q}_x = \frac{T_1 - \bar{T}_2}{R_1}$ in (22) we obtain the following evolution equation

$$\frac{\tau_S}{R_2} \frac{d\bar{T}_2}{dt} = -\frac{\bar{T}_2 - T_2^\infty}{R_2} + \frac{T_1 - \bar{T}_2}{R_1} \tag{24}$$

where

$$\tau_S = \frac{(1+\alpha) c_{\text{Si}} \ell^2}{\kappa_{\text{Si}}} \tag{25}$$

and

$$R_2 = \frac{B\ell^2}{\kappa_{\text{Si}} L} \tag{26}$$

is the viscous resistance.

Consider now the energy conservation in the heater $c_{\text{Ni}} \frac{\partial T_1}{\partial t} = -\nabla \cdot \vec{q}$. By performing volume integration of this equation with using the Kapitza interface boundary condition and the insulation condition for the other boundaries, we obtain an independent evolution equation

$$c_{\text{Ni}} h \frac{dT_1}{dt} = -\frac{T_1 - \bar{T}_2}{R_1}. \tag{27}$$

Notice that, in the derivation of the evolution equations (24,27), a simplified Kapitza interface condition have been used i.e. the hydrodynamic contributions to the interface boundary condition (10) have been neglected for simplicity. The inclusion of the hydrodynamic contributions to the boundary condition (which cause only small deviations for extremely small $L < 50$nm) can be found in the ending notes of this section.

The heater temperature $T_1$ can be obtained from the system of partial differential equations (24,27):

$$T_1 - T_2^\infty = a_1 \exp(-t/\tau_1) + a_2 \exp(-t/\tau_2). \tag{28}$$

In the case of isolated heaters ($R_1 < R_2$), the system (24,27) can be simplified and we obtain

$$\tau_1 = R_1 \frac{c_{\text{Ni}} h \tau_S / R_2}{\tau_S / R_2 + c_{\text{Ni}} h} \equiv R_1 \frac{C_1 C_2}{C_1 + C_2} \equiv R_1 C_{eq} \tag{29}$$

$$\tau_2 = \frac{B c_{\text{Ni}} \ell^2}{\kappa_{\text{Si}}} \frac{h}{L} + \frac{c_{\text{Si}} \ell^2 (1+\alpha)}{\kappa_{\text{Si}}} = c_{\text{Ni}} h R_2 + \tau_S \equiv (C_1 + C_2) R_2 \tag{30}$$

being $C_1 = c_{\text{Ni}} h$, $C_2 = \tau_S / R_2$ the heat capacities of the heater and the dam region, respectively. With these definitions, the system (24,27) is equivalent to (2) in the main text.

In the case considered here ($\tau_1 < \tau_2$ i.e. $R_1 < R_2$),

$$a_1 = \frac{\tau_S/R_2}{c_{\text{Ni}}h + \tau_S/R_2} = \frac{C_2}{C_1 + C_2} \tag{31}$$

$$a_2 = \frac{c_{\text{Ni}}h}{c_{\text{Ni}}h + \tau_S/R_2} = \frac{C_1}{C_1 + C_2} \tag{32}$$

If $R_1 \sim R_2$, then the system (24,27) needs to be solved with no approximations. This is the case of interest for small $L$ and $P = 4L$ (close-packed situation) in which we use a reduced non-local length $\ell_{\text{eff}}$ (reduced $R_2$). See the ending notes of this section for details.

For illustration, in Figure 7 we show the comparison between the analytical prediction (28) for the heater temperature evolution $T_1(t)$ with $L = 20nm$ and $P = 800nm$ compared with the Finite Elements calculation.

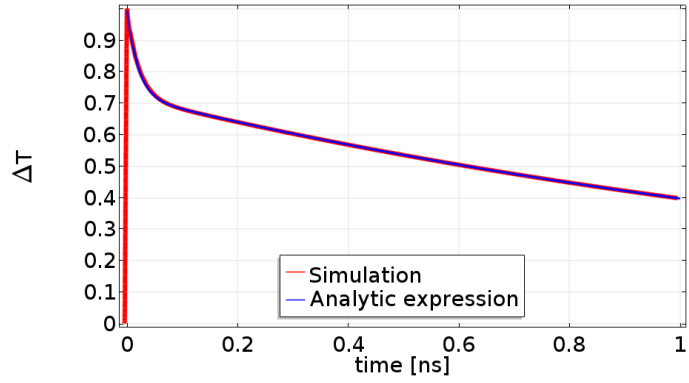

**Figure 7:** Expression (28) with parameters (29,30,31,32) compared with the simulated thermal decay of the heater according to KCM obtained using COMSOL Multiphysics for $L = 20nm$ and $P = 800nm$.

**Note I: Solutions of the system of partial differential equations (24,27)**

Here we revisit the solutions of the system of partial differential equations (24,27). Exponential solutions $\exp(tw)$ satisfy

$$w^2 + w\left(\frac{1}{R_1 C_{eq}} + \frac{1}{\tau_S}\right) + \frac{1}{\tau_S C_1 R_1} = 0 \tag{33}$$

being $w$ the roots of the system characteristic polynomial. There are always two real negative roots $w_1 = -1/\tau_1$ and $w_2 = -1/\tau_2$. In order to find the expressions (29,30) for the characteristic times we assumed $R_2 > R_1$ and hence we simplified the previous equation to

$$w^2 + w\left(\frac{1}{R_1 C_{eq}}\right) + \frac{1}{\tau_S C_1 R_1} = 0 \tag{34}$$

Regarding the weights of the double exponential, in general we have

$$a_1 = \frac{\tau_1}{C_1 R_1} \frac{\tau_2 - C_1 R_1}{\tau_2 - \tau_1} \tag{35}$$

$$a_2 = \frac{\tau_2}{C_1 R_1} \frac{C_1 R_1 - \tau_1}{\tau_2 - \tau_1} \tag{36}$$

which in the case $\tau_1 < \tau_2$ simplify to equations (31,32).

Notice that equations (33,35,36) are the general system of equations for the parameters $\tau_1, \tau_2, a_1, a_2$. Now consider the two geometrical regimes:

**(i)** Isolated situation ($P - L > 2\ell$): In this case $R_2 > R_1$ so that the approximated equation (34) can be used and the explicit analytical expressions (29,30) for $\tau_1, \tau_2$ are very close to the exact solutions of the general equation (33).

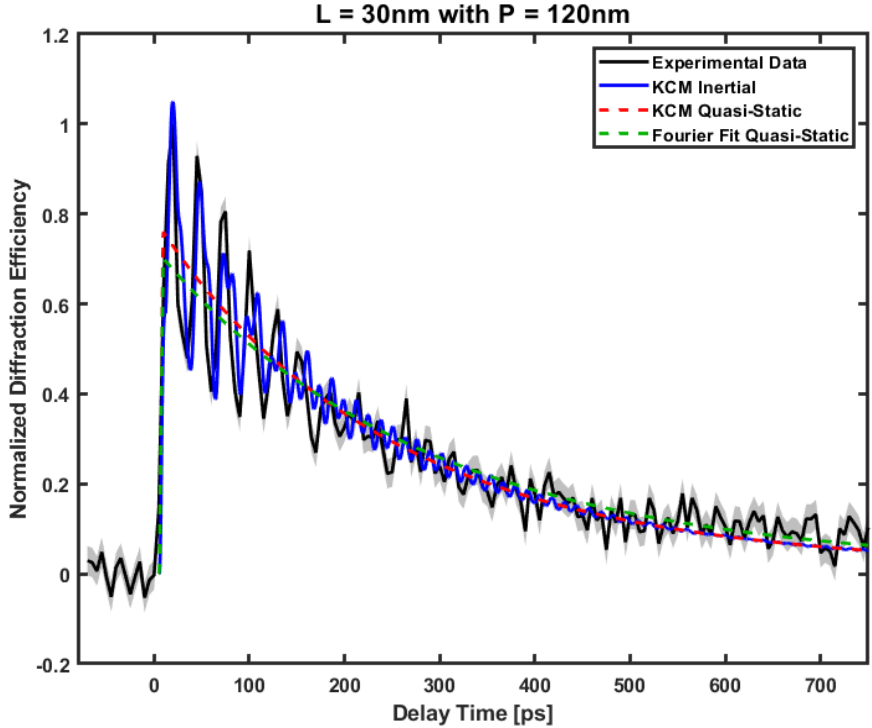

**Figure 8:** Comparison of the change in diffraction efficiency for experimental data, KCM, and an effective Fourier model for small heat source size and spacing (close-packed). The experimental change in diffraction efficiency for $L = 30$nm with $P = 120$nm is shown in black and error is embodied in grey shading. The KCM prediction is shown in blue including the full inertial calculations (solid) and the quasi-static (dashed). An effective Fourier model with a fitted thermal boundary resistance is shown in green. As predicted by the two-box model, we recover a Fourier-like behavior in this extreme close-packed situation.

**(ii)** Close-packed situation ($P = 4L$ and small $L$): In this case $\ell_{\text{eff}} = (P - L)/2$ so that $R_1 \sim R_2$. In this case the use of the approximated equation (34) is not acceptable and hence we use the exact solutions of the general equation (33) to compare with experiments. It is easy to show that by reducing $L$ with $P = 4L$, $a_2$ goes to 1 and $a_1$ goes to 0; $\tau_2$ goes to the Fourier decay time $\tau_F = c_{\text{Ni}} h R_1$ and $\tau_1$ tends to zero. Therefore, we recover Fourier in this limit and we don't expect a clearly observable double exponential decay. In particular, for $L < 50$nm, $P = 4L$ we expect a single exponential decay as in the Fourier-based description. This limit is consistent with experimental observations as shown in Figure 8.

**Note II: Inclusion of the hydrodynamic contributions to the interface boundary condition (10) in the two box model.**

Consider the hydrodynamic interface boundary condition (10) instead of the simplified Kapitza condition. We average it along the interface with expressing $\nabla \cdot \vec{q}$ and $\nabla \vec{q}$ as quantities proportional to the interface normal heat flux $\bar{q}_x$ (this is possible because of the saturation of the heat flux profile close to the interface):

$$(\bar{T}_1 - \bar{T}_2) = \bar{q}_x [R_1 - \frac{1}{\gamma^{\text{Si}} L}(\beta b_1 - \chi_{xx} b_2 - \chi_{yy} b_3)] \tag{37}$$

where

$$b_1 = \frac{\int_{-L/2}^{L/2} dy \nabla \cdot \vec{q}}{\bar{q}_x} = 1.5 \tag{38}$$

$$b_2 = \frac{\int_{-L/2}^{L/2} dy \frac{\partial q_x}{\partial x}}{\bar{q}_x} = -1 \tag{39}$$

$$b_3 = \frac{\int_{-L/2}^{L/2} dy \frac{\partial q_y}{\partial y}}{\bar{q}_x} = 2.5 \tag{40}$$

are dimensionless parameters estimated from the corresponding saturated value in the COMSOL simulations.

Now we can input this averaged boundary condition to the $\bar{T}_2$ evolution equation (22) and the heater energy conservation equation to obtain the two box model system of equations. This system is exactly the same as the one previously obtained (24,27) with a redefined interface boundary resistance $R'_1$:

$$R'_1 = R_1 + \frac{1}{\gamma^{\text{Si}} L}(-\beta b_1 + \chi_{xx} b_2 + \chi_{yy} b_3). \tag{41}$$

Therefore, by including the hydrodynamic contributions to the boundary condition we obtain a larger thermal resistance which become explicitly geometry dependent. This correction slightly increase the $\tau_1$ value for small sizes reported in Figure 6 of the main text with respect to the value obtained using the simplified Kapitza boundary resistance. However, for $L > 50$nm, we have $R_1 \sim R'_1$.

# 3 Matrix Pencil Method

We utilize the matrix pencil method (MPM) to filter the elastic wave oscillations out of the experimental change in diffraction efficiency signal. Our MPM algorithm begins by forming the time-lag co-variance matrix from an experimental time signal (i.e. the diffraction efficiency signal as a function of pump-probe time delay). We then perform the singular value decomposition of this matrix and create a scree plot of the singular values. From this plot, we separate the relevant time-correlated patterns in the data from the random noise. Next, we compute the complex exponential ($e^{(\alpha+i\beta)t}$) that best represents each relevant component. The result is a decomposition of the experimental signal into a few complex exponentials and random noise shown in Figure 9. We remove the oscillatory exponentials (exponentials with complex arguments) since they are not of interest in this study, then sum the remaining exponentials and noise. The final result is an experimental time signal without oscillations from elastic waves also shown in Figure 9. In essence, MPM is similar to a robust least squares fitting of multiple complex exponentials to the experimental data. In the presence of noise, MPM can more precisely and accurately extract the damped oscillations than a Fourier transform. A simple Fourier transform (which assumes stationary oscillations) is a sufficient choice for higher signal-to-noise, long lived periodic signals. A simple example is shown in Figure 10. Mathematical details of MPM along with validation and comparison to a Fourier transform can be found in [13, 17, 5]. An example application can be found in [10] and the numerical algorithm used in this work can be found in [9].

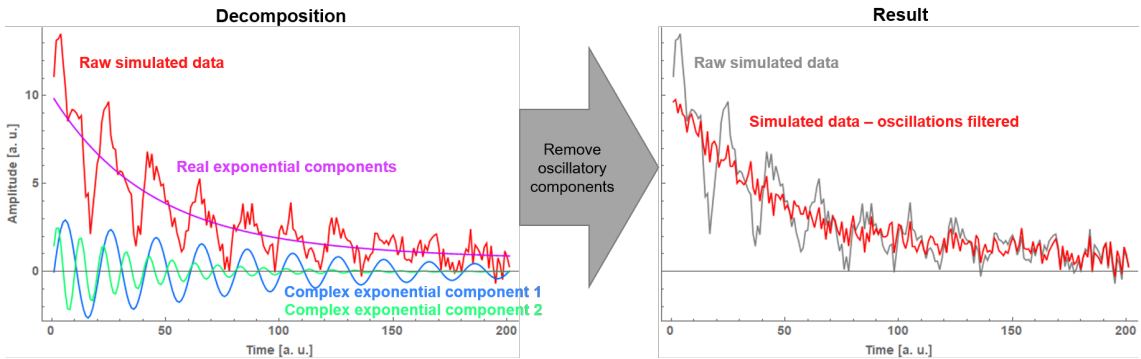

**Figure 9:** Matrix pencil method (MPM) is used to remove the oscillations of the elastic waves from the experimental data. Left is the MPM decomposition of simulated data with gaussian noise. MPM breaks down the signal into exponentials with real arguments, exponentials with complex arguments, and noise. Right is the re-constructed signal without the oscillatory (exponentials with complex arguments) components.

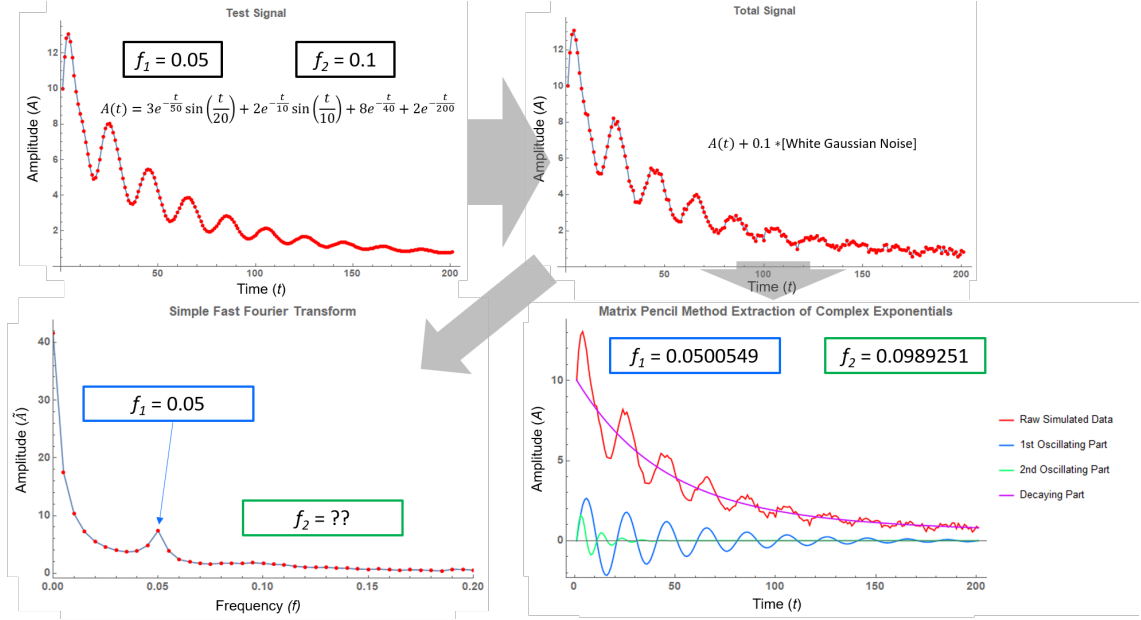

**Figure 10:** Top left shows a test signal generated by the addition of two damped sinusoids and two exponential decays (a good approximation of our experimental signal). Top right shows a small amount of white Gaussian noise added to the test signal. Bottom right shows that matrix pencil method (MPM) decomposition of the signal which identifies two oscillations. MPM extracts not only frequency values very close to the original test signal but also the decay times of each oscillation. Bottom left shows that a simple fast Fourier transform (FFT) identifies—with high accuracy—the obvious frequency but that highly damped oscillation is hidden by the noise. These results highlight the comparison of MPM and a simple FFT.

# 4 Atomic Force Microscopy

We perform atomic force microscopy (AFM) measurements to characterize the sizes and periods of the nanostructured transducers arrays (nano-gratings) utilized in the experiment. As mentioned in *Methods*, we fabricate metallic nickel nanostructures on the surface of a silicon substrate using e-beam lithography techniques. We characterize the linewidth $L$, periodicity $P$, and height $h$ of the nano-gratings for both nanoline (1D-confined) and nanodot (2D-confined) geometries. An example AFM measurement is shown in Figure 11.

We calculate $L$ by measuring the width at the top of the nanostructure in the AFM image, $P$ by measuring the distance between peaks in the auto-correlation of the AFM image, and $h$ by computing the difference between the two peaks which appear in a histogram of the heights in the AFM image (one peak corresponds to the substrate while the other corresponds to the top of the nanostructures). The results of these measurements, see Table 2, confirm the high quality of the fabrication. We use these values in the numerical KCM simulations. Because the fabrication technique creates a uniform height of nanostructures across the entire sample, we use the same average height value for all nano-gratings.

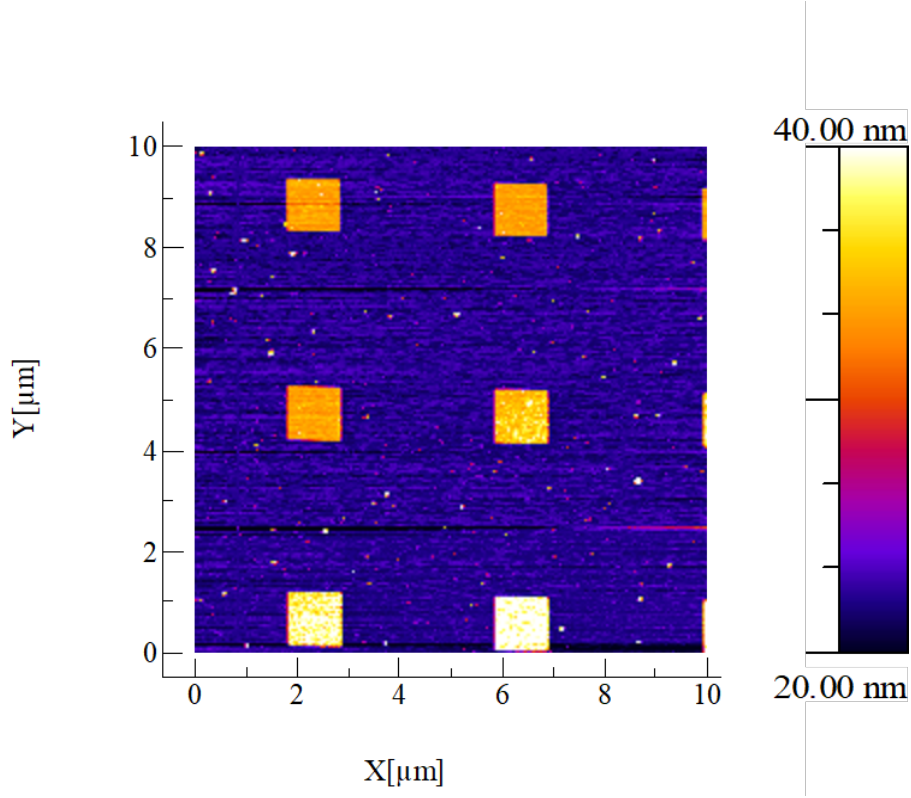

**Figure 11:** Example atomic force microscopy (AFM) image for a nominal $L = 1000nm$ and $P = 4000nm$ nanodot array

For any nano-grating geometries with no $L$ or $P$ AFM values in Table 2, we use nominal $L$ and $P$ with the average value for $h$ in the numerical KCM simulations.

## 5  More Details on Extraction of Decay Times

In Figure 6 of the article, the agreement between numerical predictions and experimental measurements for $\tau_1$ and $\tau_2$ is quite good. However, there exists disagreement between numeric simulations and experiments for the interface decay ($\tau_1$) for small linewidth $L$ and for the hydrodynamic decay ($\tau_2$) for large linewidth $L$. For small $L$, the amplitude of the interface decay ($\tau_1$) is much smaller than the amplitude of the hydrodynamic decay ($\tau_2$). Thus, the experimental extraction of $\tau_1$ becomes more challenging. For large $L$, the interface decay dominates over the hydrodynamic decay making it more difficult to experimentally observe the hydrodynamic decay.

When extracting $\tau_1$ from the experimental measurements for small $L$, we encounter additional challenges since not only is the amplitude of the $\tau_1$ exponential small but also $\tau_1$ is small. When $\tau_1$ is small, it is on a similar scale as the finite response time of the nanostructures and an elastic wave oscillation. Although the pump laser pulse has an ultrashort duration ($\approx 25\text{fs}$), the deformation of the

|  | Nanoline arrays | | | | Nanodot arrays | | |
|---|---|---|---|---|---|---|---|
| Nominal L,P [nm] | AFM L [nm] | AFM P [nm] | AFM h [nm] | Nominal L,P [nm] | AFM L [nm] | AFM P [nm] | AFM h [nm] |
| 1000,4000 | 950±40 | 4050±40 | 11.5±1 | 1000,4000 | 1000±20 | 4080±40 | 11.5±1 |
| 300,1200 | 260±10 | 1240±10 | 11.5±1 | 300,1200 | 300±20 | 1220±20 | 11.5±1 |
| 200,800 | - | - | 11.5±1 | 50,400 | 51±4 | 406±4 | 11.5±1 |
| 100,400 | - | - | 11.5±1 | 50,200 | 54±5 | 200±5 | 11.5±1 |
| 50,400 | 51±3 | 405±4 | 11.5±1 | 30,120 | - | - | 11.5±1 |
| 50,200 | 49±2 | 205±4 | 11.5±1 | - | - | - | - |
| 30,400 | 35±3 | 407±3 | 11.5±1 | - | - | - | - |
| 30,120 | 30±3 | 122±2 | 11.5±1 | - | - | - | - |
| 20,400 | 29±4 | 406±5 | 11.5±1 | - | - | - | - |
| 20,80 | - | - | 11.5±1 | - | - | - | - |

**Table 2:** Table of compiled AFM measurements of nano-grating geometries. $L$ is linewidth, $P$ is period, and $h$ is height. Uncertainty is calculated by standard deviation of measurements and AFM image pixel size. We use WSxM to analyze the raw AFM images [11].

system responds on the time scale of 10ps (we measure surface deformation with the probe beam). These inertial effects can inhibit our ability to extract $\tau_1$

It is challenging to extract $\tau_1$ for small $L$, close-packed (small period $P$) nanostructure arrays. In the extreme case where $L = 20$nm in the close-packed regime, the predicted $\tau_1 \approx 10$ps while the experimentally observed time required for the nanostructure to reach maximum thermal expansion is $\approx 15$ps. Therefore, we cannot extract $\tau_1$ for very small, close-packed nanostructure arrays. For slightly larger $L$, such as $L = 60$nm, we can extract $\tau_1$; however, the oscillations from elastic waves, the finite response time of the surface deformation, and the chosen experimental time sampling can artificially increase the extracted value of $\tau_1$. In the example shown in Figure 12, the $\tau_1$ extracted from experimental data is much different than that predicted by quasi-static numerical simulations. However, we also find deviations in the value extracted from inertial numerical simulations compared to the original value predicted by numerical quasi-static simulation, even though the both simulations are in agreement as shown in the article. Since $\tau_1$ is on the similar time scale as one elastic wave oscillation, we cannot extract $\tau_1$ with high accuracy. In Figure 12, sampling the numerical inertial simulation to match the experimental time resolution and then filtering the elastic wave oscillations does not result in a more accurate value. This example suggests that inertial effects, e.g. elastic waves, can inhibit the ability to observe $\tau_1$ in this regime.

It is also challenging to extract $\tau_1$ for small $L$, effectively isolated (large period $P$) nanostructure arrays. Because the change in diffraction efficiency signal is dominated by nanostructure expansion at small $L$ [6], the signal-to-noise ratio is related to the duty cycle, i.e. the ratio $L/P$. For small $L$ and large $P$, the signal-to-noise ratio is low and thus extraction of $\tau_1$ is challenging. As shown

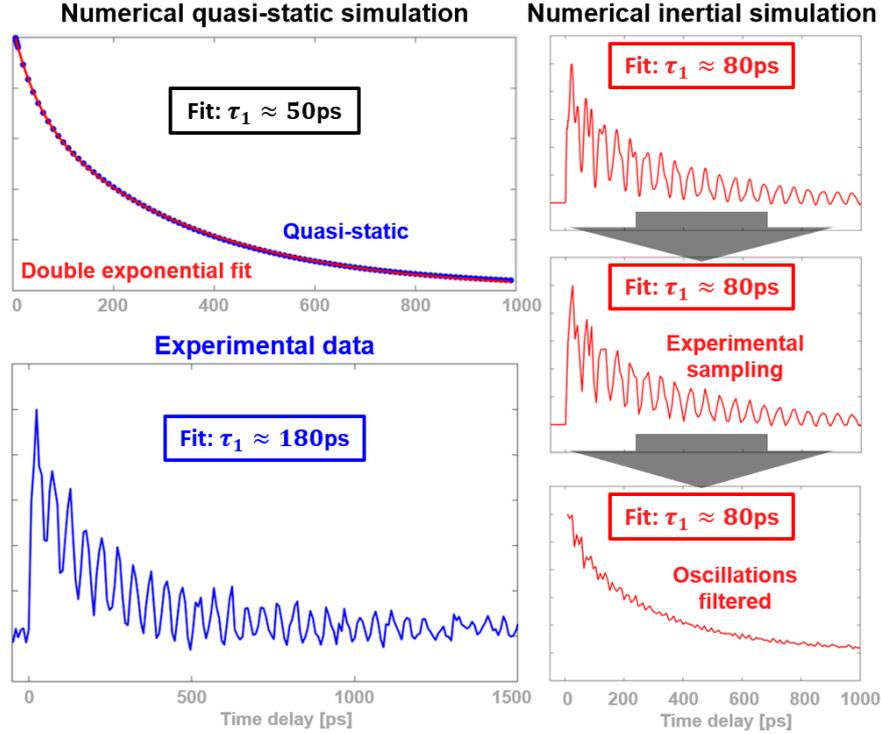

**Figure 12:** Example of inertial effects inhibiting observation of $\tau_1$ where linewidth $L = 60$nm and period $P = 240$nm. Extracting $\tau_1$ from numerical quasi-static simulations results in a value of $\approx 80$ps as seen in upper left. This $\tau_1$ value is much different than what we experimentally observe, shown in the lower left. However, the values of $\tau_1$ extracted from a numerical inertial simulation are also larger than predicted by quasi-static simulations, shown in the top right. This example suggests that inertial effects, e.g. elastic waves, can inhibit the ability to observe $\tau_1$ in this regime. With data sampled at the experimental resolution, filtering the acoustic oscillations does not result in a more accurate value for $\tau_1$ shown in bottom right.

in Figure 13, the $\tau_1$ extracted from experimental data is almost 4 times higher than that predicted by quasi-static numerical simulations. The extraction of $\tau_1$ from the numerical inertial simulation is about 30% higher than the quasi-static prediction. This discrepancy between the two simulations is most likely due to inertial effects as previously described. However, if we sample the inertial simulation at the experimental time resolution, the extracted $\tau_1$ is now three times higher than the quasi-static prediction. Although our experimental capabilities allow us to capture $< 10$fs dynamics, observing the entire thermal decay ($\sim$ ns) with fs resolution creates massive experimental data sets. Therefore, we chose the experimental time sampling to be on the order of $10$ps, even though it is far from technique limits. If we add Gaussian noise (same signal-to-noise ratio as experiment) to the inertial simulation sampled at experimental resolution, our fit algorithm cannot obtain reasonable values for a fitted $\tau_1$. These results suggests that inertial effects, sampling, and noise inhibit our ability to accurately

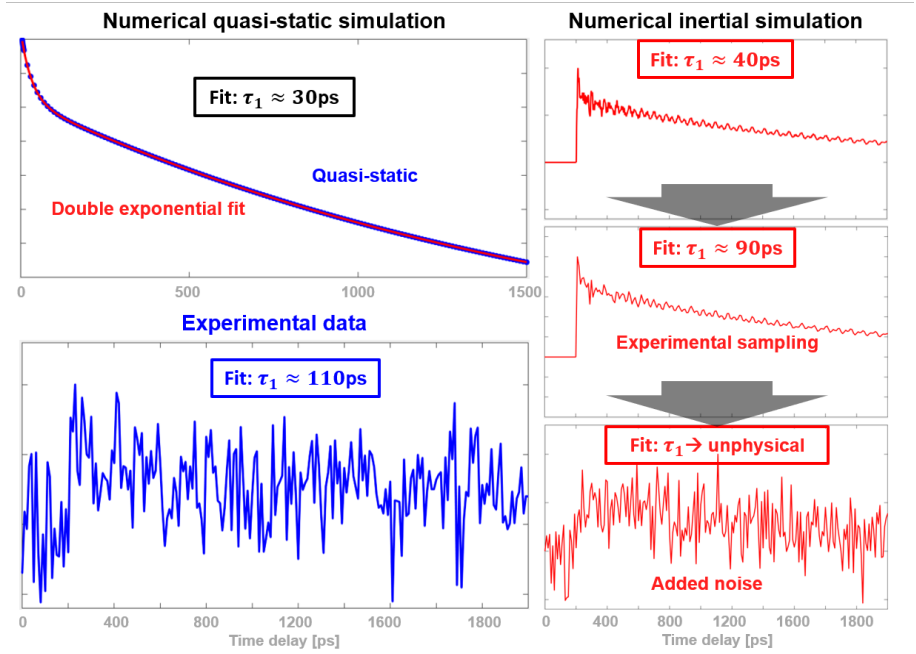

**Figure 13:** Example of inertial effects, sampling, and noise inhibiting observation of $\tau_1$ where linewidth $L = 20$nm and period $P = 400$nm. The value of $\tau_1$ extracted from experimental data (bottom left) is almost 4 times higher than that predicted by numerical quasi-static simulations (top left). The extracted values of $\tau_1$ from the numerical inertial simulations is about 30% higher than the quasi-static prediction (top right). However, if we sample the inertial simulation at the experimental time resolution and add Gaussian noise, we cannot extract a reasonable value for $\tau_1$ (bottom right). Experimental data is from [6]

extract $\tau_1$ for extremely small $L$. Note, that the error bars in Figure 6 in the article are computed from the standard deviation of the extracted $\tau_1$ values; therefore, the error bars embody the precision of the experimental data and not necessarily the accuracy of the fit procedure. Note that the quasi-static predictions shown in Figures 12 and 13 use different intrinsic parameters than the results in the article.

# References

[1] F X Alvarez, D Jou, and A Sellitto. "Phonon hydrodynamics and phonon-boundary scattering in nanosystems". In: *Journal of Applied Physics* 105.1 (2009), p. 14317.

[2] F. Banfi et al. "Ab initio thermodynamics calculation of all-optical time-resolved calorimetry of nanosize systems: Evidence of nanosecond decoupling of electron and phonon temperatures". In: *Physical Review B - Condensed Matter and Materials Physics* 81 (15 2010), pp. 1–5. ISSN: 10980121.

[3] A. Beardo et al. "Hydrodynamic Heat Transport in Compact and Holey Silicon Thin Films". In: *Phys. Rev. Applied* 11 (3 Mar. 2019), p. 034003.

[4] A. Beardo et al. "Phonon hydrodynamics in frequency-domain thermoreflectance experiments". In: *Phys. Rev. B* 101 (7 Feb. 2020), p. 075303.

[5] J. E. Fernandez del Rio and T. K. Sarkar. "Comparison between the matrix pencil method and the Fourier transform technique for high-resolution spectral estimation". In: *Digit. Signal Process.* 6 (1996), pp. 108–125.

[6] Travis D. Frazer et al. "Engineering Nanoscale Thermal Transport: Size- and Spacing-Dependent Cooling of Nanostructures". In: *Phys. Rev. Applied* 11 (2 Feb. 2019), p. 024042.

[7] Yangyu Guo and Moran Wang. "Phonon hydrodynamics for nanoscale heat transport at ordinary temperatures". In: *Physical Review B* 97.3 (2018). ISSN: 24699969.

[8] R. A. Guyer and J. A. Krumhansl. "Solution of the Linearized Phonon Boltzmann Equation". In: *Physical Review* 148.2 (Aug. 1966), pp. 766–778. ISSN: 0031-899X.

[9] K. M. Hoogeboom-Pot. "Uncovering new thermal and mechanical behavior at the nanoscale using coherent exterme ultraviolet light". PhD thesis. CU Boulder, 2015.

[10] K. M. Hoogeboom-Pot et al. "Nondestructive measurement of the evolution of layer-specific mechanical properties in sub-10 nm bilayer films". In: *Nano Letters* 16 (2016), pp. 4773–4778.

[11] I. Horcas et al. "WSXM: A software for scanning probe microscopy and a tool for nanotechnology". In: *Rev. Sci. Instrum.* 78 (2007), p. 013705.

[12] Yongjie Hu et al. "Spectral mapping of thermal conductivity through nanoscale ballistic transport". In: *Nature Nanotechnology* 10.8 (June 2015), pp. 701–706. ISSN: 1748-3387.

[13] Y. Hua and T. K. Sarkar. "Matrix pencil method to estimate the parameters of exponentially damped/undamped sinusoids in noise". In: *IEEE Trans. Acoust., Speeh, Signal Process.* 38 (1990), pp. 814–824.

[14] M. I. Kaganov, I. M. Lifshitz, and L. V. Tanatarov. "Relaxation between Electrons and the Crystalline Lattice". In: *Soviet Physics Journal of Experimental and Theoretical Physics* 4 (2 1957), pp. 173–178.

[15] L. D. Landau and Lifshitz. " Theory of Elasticity: Vol. 7 of Course of Theoretical Physics ". In: (1960).

[16] Damiano Nardi et al. "Probing Thermomechanics at the Nanoscale: Impulsively Excited Pseudosurface Acoustic Waves in Hypersonic Phononic Crystals". In: *Nano Letters* 11.10 (2011). PMID: 21910426, pp. 4126–4133. eprint: https://doi.org/10.1021/nl201863n.

[17] T. K. Sarkar and O. Pereira. "Using the matrix pencil method to estimate the parameters of a sum of complex exponentials". In: *IEEE Antennas and Propagation Mag.* 37 (1995), pp. 48–55.

[18] P. Torres et al. "Emergence of hydrodynamic heat transport in semiconductors at the nanoscale". In: *Physical Review Materials* 2.7 (2018), p. 076001. ISSN: 2475-9953.

[19] P. Torres et al. "First principles kinetic-collective thermal conductivity of semiconductors". In: *Physical Review B* 95.16 (Apr. 2017), p. 165407. ISSN: 2469-9950. arXiv: 1606.01149.

[20] R Zhu et al. "Atomistic calculation of elastic moduli in strained silicon". In: *Semiconductor Science and Technology* 21.7 (June 2006), pp. 906–911.

[21] Amirkoushyar Ziabari et al. "Full-field thermal imaging of quasiballistic crosstalk reduction in nanoscale devices". In: *Nature Communications* 9.1 (Dec. 2018), p. 255. ISSN: 2041-1723.